\documentclass[aps,prl,reprint,amsmath,amssymb,superscriptaddress,floatfix]{revtex4-1}

\usepackage{natbib} 
\usepackage[colorlinks,
            linkcolor=blue,
            anchorcolor=blue,
            citecolor=blue,
urlcolor=blue
            ]{hyperref}
\usepackage{graphicx}
\usepackage[FIGTOPCAP,raggedright,nooneline]{subfigure}
\usepackage{braket}
\usepackage{mathtools}
\usepackage{float}
\usepackage{longtable}
\begin{document}

\title{Experimental test of contextuality in quantum and classical systems}

\author{Aonan Zhang}
\affiliation{National Laboratory of Solid State Microstructures, College of Engineering and Applied Sciences and School of Physics, Nanjing University, Nanjing 210093, China}
\affiliation{Collaborative Innovation Center of Advanced Microstructures, Nanjing University, Nanjing 210093, China}
\author{Huichao Xu}
\affiliation{National Laboratory of Solid State Microstructures, College of Engineering and Applied Sciences and School of Physics, Nanjing University, Nanjing 210093, China}
\affiliation{Collaborative Innovation Center of Advanced Microstructures, Nanjing University, Nanjing 210093, China}
\author{Jie Xie}
\affiliation{National Laboratory of Solid State Microstructures, College of Engineering and Applied Sciences and School of Physics, Nanjing University, Nanjing 210093, China}
\affiliation{Collaborative Innovation Center of Advanced Microstructures, Nanjing University, Nanjing 210093, China}
\author{Han Zhang}
\affiliation{National Laboratory of Solid State Microstructures, College of Engineering and Applied Sciences and School of Physics, Nanjing University, Nanjing 210093, China}
\affiliation{Collaborative Innovation Center of Advanced Microstructures, Nanjing University, Nanjing 210093, China}
\author{Brian J. Smith}
\email[]{bjsmith@uoregon.edu}
\affiliation{Department of Physics and Oregon Center for Optical, Molecular, and Quantum Science, University of Oregon, Eugene, Oregon 97403, USA}
\author{M. S. Kim}
\affiliation{QOLS, Blackett Laboratory, Imperial College London, London SW7 2AZ, United Kingdom}
\affiliation{Korea Institute for Advanced Study, Seoul, 02455, Korea}
\author{Lijian Zhang}
\email[]{lijian.zhang@nju.edu.cn}
\affiliation{National Laboratory of Solid State Microstructures, College of Engineering and Applied Sciences and School of Physics, Nanjing University, Nanjing 210093, China}
\affiliation{Collaborative Innovation Center of Advanced Microstructures, Nanjing University, Nanjing 210093, China}


\date{\today}

\begin{abstract}
Contextuality is considered as an intrinsic signature of non-classicality, and a crucial resource for achieving unique advantages of quantum information processing. However, recently there have been debates on whether classical fields may also demonstrate contextuality. Here we experimentally configure a contextuality test for optical fields, adopting various definitions of measurement events, and analyse how the definitions affect the emergence of non-classical correlations. The heralded single photon state, a typical non-classical light field, manifests contextuality in our setup, while contextuality for classical coherent fields strongly depends on the specific definition of measurement events which is equivalent to filtering the non-classical component of the input state. Our results highlight the importance of definition of measurement events to demonstrate contextuality, and link the contextual correlations to non-classicality defined by quasi-probabilities in phase space.
\end{abstract}

\pacs{}

\maketitle

\par
The probabilistic nature of measurements in quantum theory, in which observables do not have pre-determined values, is one of the most distinguishing features in contrast to classical physics. This property is accentuated when considering measurements of compatible observables and can be formulated as contextuality~\cite{Kochen1967,JOHNS.1966}, which means that the outcome of a measurement will depend on the context of the other measurements being performed. Contextuality can be demonstrated through the violation of non-contextual (NC) inequalities usually involving dichotomic measurements with binary outcomes. Similar to non-locality originated from quantum entanglement, contextuality is often considered as a signature of non-classical behavior~\cite{Lapkiewicz_2011,Kirchmair2009,PhysRevLett.109.150401,PhysRevLett.90.250401,PhysRevLett.117.170403}, and a critical resource of quantum computation \cite{Howard2014,PhysRevX.5.021003,PhysRevLett.115.030501,PhysRevA.88.022322,PhysRevLett.119.120505} and quantum communication \cite{PhysRevA.72.012325}.
\par
Recently, there have been a new type of investigation called classical entanglement, which draw the analogy between non-separability among different degrees of freedom of classical light fields and entanglement among different particles, and suggest that features such as entanglement are not unique to quantum theory~\cite{Spreeuw1998,T_ppel_2014,Karimi_2015,Qian_2015,Kagalwala_2012}. As a natural extension of classical entanglement, contextuality in classical systems has also drawn much interest. In particular, there have been debates on whether classical fields, \textit{i.e.} coherent or stochastic light beams, can reproduce the correlations to violate NC inequalities~\cite{PhysRevLett.116.250404,Li2017} or not~\cite{Markiewicz2019}, and whether such correlations may enhance the performance of certain applications~\cite{Berg-Johansen15,Diego2016,Ndagano2017}.
\par
To date, tests of contextuality and entanglement in classical systems have used measurement devices different from those used for quantum systems~\cite{Kagalwala_2012,Qian_2015,PhysRevLett.116.250404}, and the binary outcomes are not the direct response of the devices but rather defined artificially with certain selection rules and through proper considerations of measurement events~\cite{PhysRevLett.116.250404}. Although these devices are elaborately designed to ensure the classical systems reproduce probability distributions given by the quantum theory, they have, nevertheless, different working mechanisms other than those used for quantum systems. This difference, together with the specific selection rules on measurement outcomes, makes the comparison between the behaviors of quantum and classical systems obscure, and the relation between contextuality and non-classicality of the system remains unspecified.

\par
To resolve the obscurity, in this work we examine the violations of a fundamental NC inequality for both single photon state and coherent states with the same setup and different measurement events. Through this unified comparison, our configuration of the test confirms contextuality of single photon state, and shows how contextual correlations emerge for coherent states under proper consideration of measurement events. Our results establish the relation between the violation of the NC bound for a linear-optical setup and the non-classicality of the measured field.
\begin{figure}
  \centering
  \includegraphics[width=\linewidth]{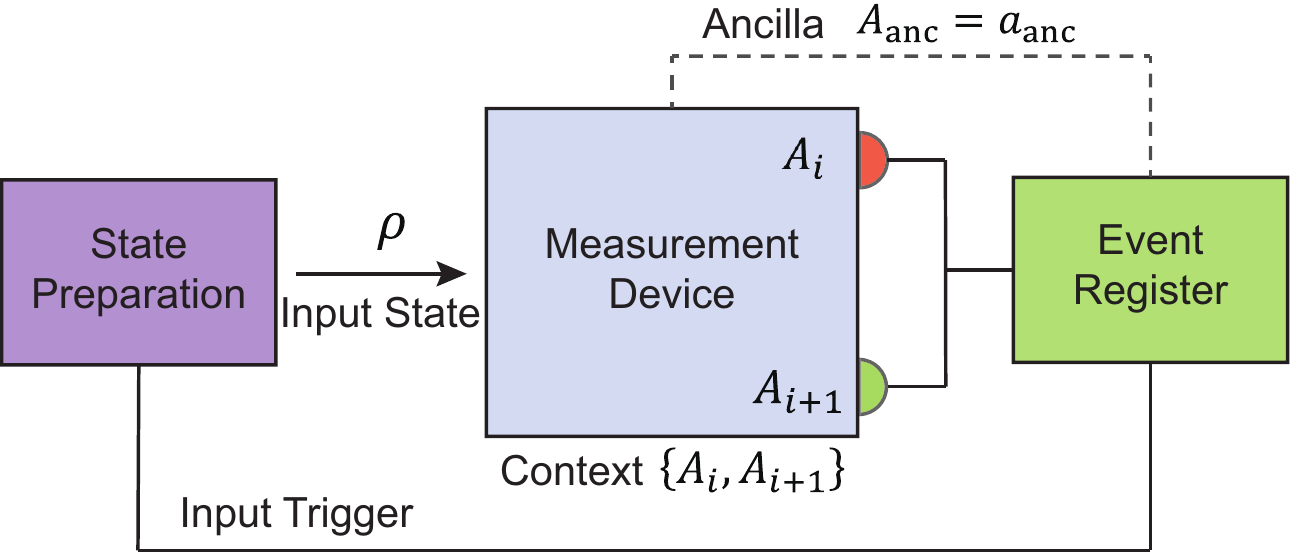}
\caption{\label{fig:framework}The scenario of testing the KCBS inequality. A physical system is prepared in state $\rho$ and sent to a measurement device, which performs measurements of compatible observables $\{A_i,A_{i+1}\}$ and produces outcomes $\{a_i=\pm1, a_{i+1}=\pm1\}$. The measurement device can be set to different configurations to realize distinct contexts $\{A_i,A_{i+1}\}$. An event register records the statistics of outcomes (including potential ancillary outcomes $A_\text{anc}=a_\text{anc}$).}
\end{figure}
\par
\textit{The scenario of testing contextuality}.---As shown in Fig. \ref{fig:framework}, a test of contextuality consists of a physical system in a given state $\rho$ as input and measurement device including a set of observables $\{A_i\}$, where $i$ labels a particular observable. In each trial, the device performs measurements of several compatible observables constituting a measurement context, for example $\{A_i,A_{i+1}\}$~\cite{PhysRevA.92.062125}, and produces the well-defined measurement outcomes $\{a_i,a_{i+1}\}$. From the operational point of view, two observables are compatible if they can be jointly measured in a single device~\cite{Kochen1967,PhysRevA.81.022121}. In actual implementations, there may be outcomes besides $A_i$ and $A_{i+1}$, which can be encapsulated in an ancillary observable $A_\text{anc}=a_\text{anc}$ ($a_\text{anc}=0,1,...,d$). 

\par
An NC inequality is usually associated with a parameter $\beta$ involving the linear combination of the probabilities $P$ of a set of events $E$ \cite{1367-2630-13-11-113036}. The Klyachko-Can-Binicio{\u{g}}lu-Shumovsky (KCBS) inequality \cite{Klyachko_2008} is the most fundamental NC inequality satisfied by all non-contextual hidden variable models \cite{PhysRevLett.110.060402}. Consider five dichotomic observables $A_i \ (i=1,2,...,5)$, each taking a value $+1$ or $-1$. Assuming the outcomes of observables reveal a pre-determined joint probability distribution, we arrive at the KCBS inequality \cite{Klyachko_2008}
\begin{equation}
\beta=\langle A_1 A_2\rangle+\langle A_2 A_3\rangle+\langle A_3 A_4\rangle+\langle A_4 A_5\rangle+\langle A_5 A_1\rangle\geq -3.
\label{eq:KCBS}
\end{equation}
Violation of this inequality implies that the tested system is contextual. Quantum theory predicts the maximum violation of the inequality to be $5-4\sqrt{5}\approx -3.944$~\cite{Klyachko_2008,PhysRevLett.110.060402,arXiv1010.2163,PhysRevLett.110.260406,PhysRevA.98.050102}.
\par
In the KCBS inequality, the correlation between $A_i$ and $A_{i+1}$, of the form $\langle A_i A_{i+1}\rangle$, can be measured by the joint probabilities of all possible outcomes
\begin{widetext}
\begin{eqnarray}
\langle A_i A_{i+1}\rangle=&&P(A_i=1,A_{i+1}=1)+P(A_i=-1,A_{i+1}=-1)-P(A_i=-1,A_{i+1}=1)-P(A_i=1,A_{i+1}=-1)\nonumber \\
=&&1+4P(A_i=-1,A_{i+1}=-1)-2P(A_i=-1)-2P(A_{i+1}=-1).
\label{eq:correlation}
\end{eqnarray}
\end{widetext}
Here $P(A_i=a_i,A_{i+1}=a_{i+1})$ denotes the probability of the event ``the outcomes $a_i,a_{i+1}$ are acquired when the measurements $A_i,A_{i+1}$ are performed''.
\begin{figure*}
  \centering
  \includegraphics[width=\linewidth]{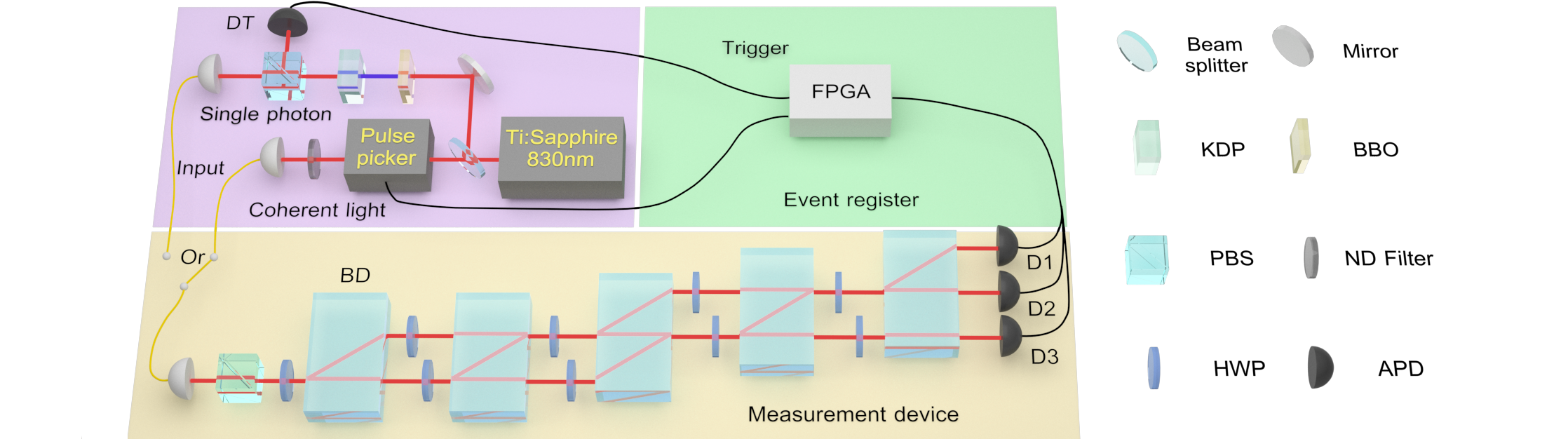}
\caption{\label{fig:setup}Experimental setup for testing the KCBS inequality with single photons and coherent light. Single photons are heralded using spontaneous parametric down conversion; the classical light is a weak coherent state generated by a Ti:Sapphire laser and attenuated by neutral density (ND) filters (purple area). The repetition rate of the pulses is reduced by a pulse picker (1.9MHz). The measurement device (yellow area) is realized by a linear optical network, comprised of birefringent beam displacers (BDs), half wave plates (HWPs) and avalanche photodiodes (APDs). A FPGA registers relevant events (green area) with a coincidence window of 4.5 ns. BBO:  $\beta$-barium borate crystal, KDP: potassium di-hydrogen phosphate crystal and PBS: polarizing beam splitter.} 
\end{figure*}
\par
\textit{Experiment.}---Figure \ref{fig:setup} illustrates the experimental setup to test the KCBS inequality using heralded single photons or classical coherent light fields. The former are generated through spontaneous parametric downconversion (SPDC) within a potassium di-hydrogen phosphate (KDP) crystal pumped by a frequency doubled pulsed Ti:sapphire laser (830nm, 76MHz)~\cite{PhysRevLett.100.133601}, while the latter is implemented with a set of coherent states generated by a Ti:sapphire laser with a reduced repetition rate (1.9MHz) using a pulse picker and under different attenuation. The SPDC is generated at a low pump regime and an avalanche photodiode (APD) in the trigger arm heralds the generation of a single photon state in the other arm as the multi-photon components are negligible. This can be confirmed by the value of the second order correlation function for heralded single photons $g^{(2)}(0)=0.0397(38)$. The detected count rate of single photons is $\sim$57000 counts/series (in 1.4s) under a heralding efficiency $\eta_H=16.2\%$ accounting for all the losses.
\par
The test of the KCBS inequality involves three modes of the light field. In our setup, the three modes are defined as one spatial mode of the light field, and the vertical polarization and horizontal polarization in another spatial mode. The light field is injected in the horizontal polarization in the second spatial mode. The measurement devices are composed of an optical network to perform transformations on polarization and spatial modes of the input optical field and three APDs to produce outcomes. We change the settings of waveplates to realize the measurements of a different pair of observables, and thereby, distinct contexts. The transformation from one context to the other leaves one of the two measurements unaffected, ensuring the measurement is physically the same in the two contexts.  The two measurements in each context are realized in a single device and defined by different detectors, thus they are co-measurable and compatible. We assign $-1$ for the measurement outcome when the corresponding APD gives a click and $+1$ when it does not. Two APDs give the outcomes of measurements $A_i,A_{i+1}$, and the third APD plays the role of the ancilla, used to identify the measurement event, i.e. only one detector clicks, at least one detector clicks, etc. The measurement of $A_i$ corresponds to a transformation on the optical modes from the input configuration $(0, 0, 1)^T$ to $c_i\left(\cos(4\pi i/5),\sin(4\pi i/5),\sqrt{\cos(\pi/5)}\right)^T$ ($c_i$ is the normalization coefficient)~\cite{Supplementary}. All the relevant twofold, threefold and fourfold coincidences between the 3 APDs and the trigger signal are registered using a field-programmable gate array (FPGA) to give the joint probabilities in $\langle A_i A_{i+1}\rangle$~\cite{Supplementary}. We record clicks of each context in 150 series, 1.4s each, to acquire the expectation value and standard uncertainty.
\par
The data is registered with respect to the \textit{measurement event}, i.e., a successful trial of the experiment, which can be defined with the aid of ancillary outcomes and the input trigger. For one single photon input, ideally one and only one detector can ``click'', neglecting dark counts, and thus provides a well-defined ``measurement event''. In contrast, when the coherent light field, which is considered to be classical, is input to the measurement device, there is a non-negligible probability for more than one detector to click due to the multi-photon component within the coherent state $|\alpha\rangle$. Moreover, there are situations that no detector clicks even a state has been input, due to the inherent vacuum component. This fact leads to an inconclusive definition of ``measurement event'' for classical inputs if we only take the measurement outcomes without referring to the input. In the following analyses, we consider three conditions to define the measurement event: (i) $E_1$: only a single detector clicks; (ii) $E_2$: at least one detector clicks; (iii) $E_3$: the full click statistics, i.e. no post-selection on detectors clicking or not clicking. Generally speaking, $E_1\subset E_2 \subset E_3$. For a perfect single photon input without any loss, $E_1=E_2=E_3$ and thus the measurement event is defined unambiguously.
\par
In the experiment the presence of single photons is heralded by the detection of photons in the trigger detector $D_T$. However, inevitable experimental imperfections, in particular optical losses and finite detection efficiencies, result in a reduced heralding efficiency and add the vacuum component to the setup, rendering the statistics of measurement outcomes different from those with the perfect single photon input. Since we are interested in the contextuality within different \textit{input} states, we resort to fair sampling assumption in the sense that the lost photons would have the same behavior as the registered photons. This assumption essentially removes the additional vacuum component and restores the result with perfect single photons, resulting in $E_1\approx E_2 = E_3$ in our experiment. Then a measurement event happens when at least one of the three detectors clicks conditioned on the trigger detector registering photons. By collecting statistics under such definition, we obtained the KCBS value $\beta=-3.9176(60)$, showing a clear violation of the KCBS inequality and confirming that the indivisible single-photon system is undoubtedly contextual. The results without fair sampling, which take into account the vacuum component caused by losses as part of the state, are given in the Supplemental Material~\cite{Supplementary}. In this case, the three definitions of measurement event do not necessarily give the same result ($E_1\approx E_2 \neq E_3$ for our source) and a heralding efficiency above $89.65\%$ is required for an unambiguous violation of the KCBS inequality.
\begin{figure*}
\centering
\addtocounter{figure}{1}
\subfigure{\includegraphics[width=0.45\linewidth]{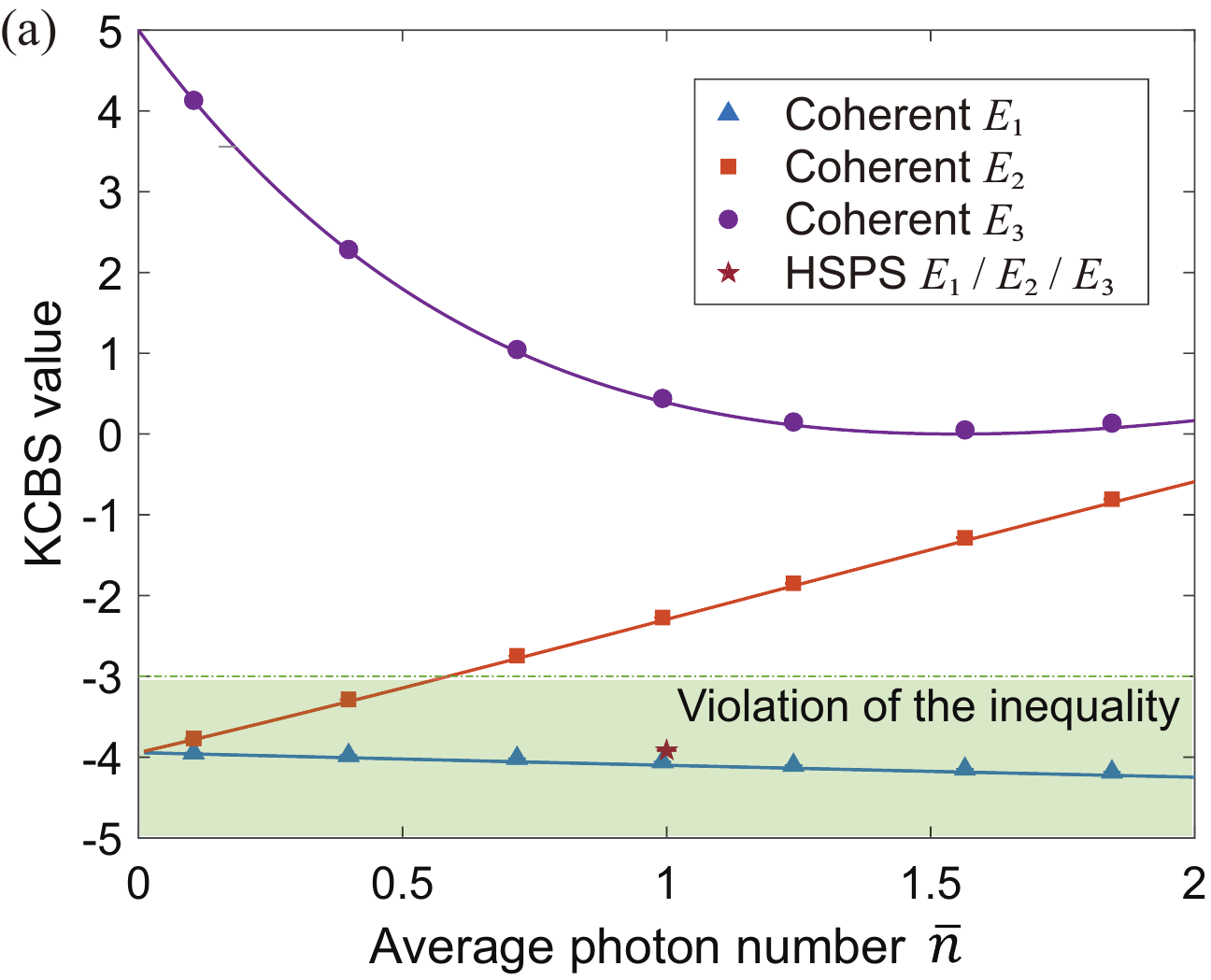}\label{fig:results:a}}
\quad
\subfigure{\includegraphics[width=0.45\linewidth]{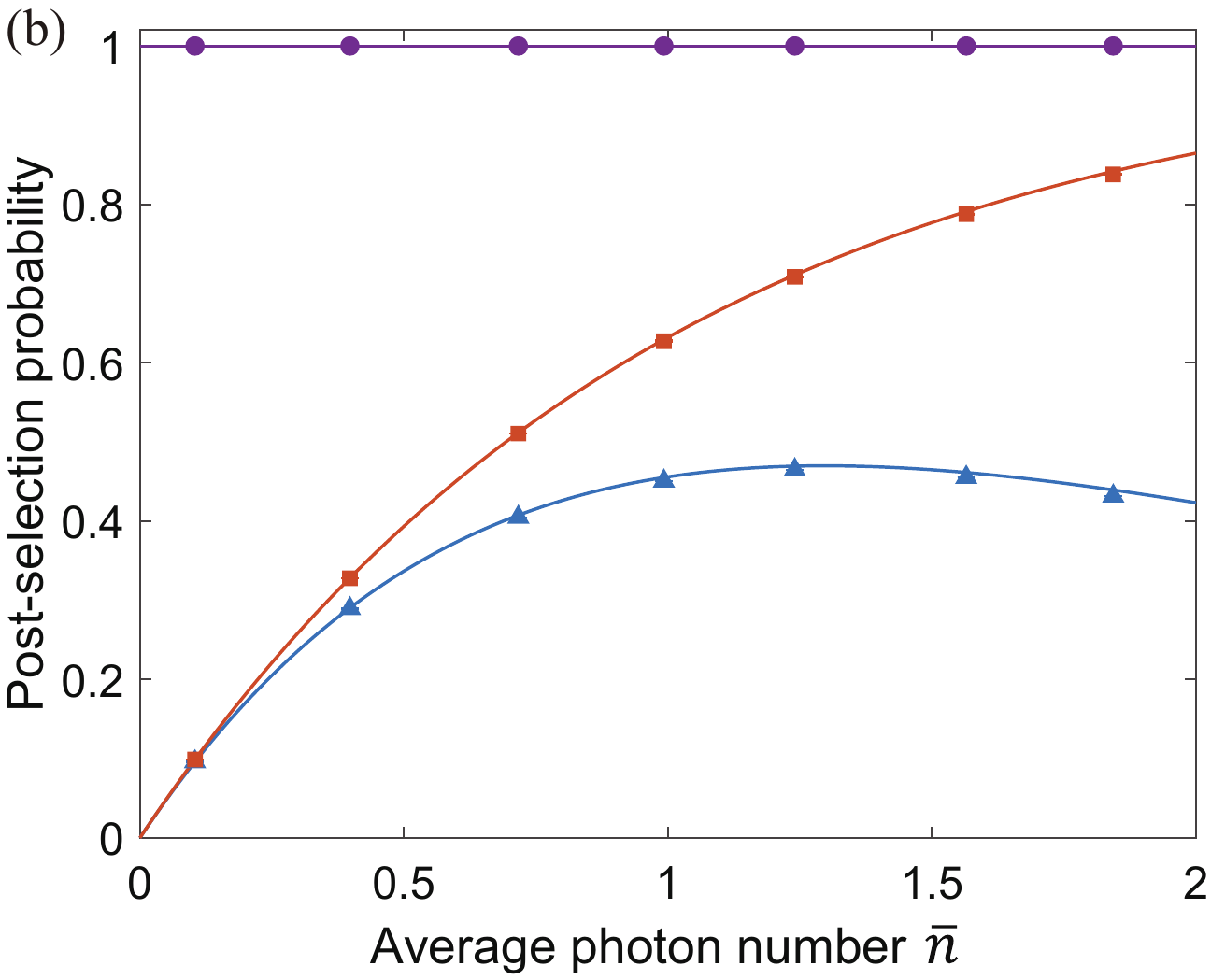}\label{fig:results:b}}
\addtocounter{figure}{-1}
\caption{\label{fig:results} (a) Experimental KCBS values (markers) and theoretical predictions (solid lines) for coherent light fields under different measurement events $E_j$: $E_1$ (triangle, blue), $E_2$ (square, red) and $E_3$ (dot, purple). The experimental result for heralded single photon source (HSPS) with fair sampling (pentagram, red) is also plotted for comparison. (b) Post-selection probabilities for coherent light fields under different $E_j$. The error bars are too small~\cite{Supplementary} to see in the two figures. }
\end{figure*}
\par
For the situation with the classical light field as the input, fair sampling cannot be applied in the same way since it is hard to distinguish between the vacuum component inherent to the input state and those appear due to loss. Yet given that the coherent state remains a coherent state after loss, we can alternatively take all the vacuum components as part of the state. In this way the losses are lumped together in the state preparation and backed out from the measurement devices. The experimental results for the classical light field are shown in Fig. \ref{fig:results:a}, where $\bar{n}=|\alpha|^2$ represents the average photon number per pulse of the coherent state. The violation of the NC inequality under $E_1$ is similar to that of single photons, which reaches the quantum bound. Note that due to the saturation of the APD, the violations can even be stronger than the quantum bound \cite{Supplementary}. $E_2$ takes a simultaneous response of different detectors into consideration, but still filters out the contribution of the vacuum component which does not fire the detectors. As a result, a significant violation is observed when the zero-click probability is considerable; and the violation disappears with the increase of $\bar{n}$ because the zero-click probability gets negligible. For $E_3$ we record all events with respect to the repetition rate of the input state. Under this condition, coherent states cannot violate the KCBS inequality.  The experimental results are in agreement with the theoretical predictions~\cite{Supplementary}. The result for the heralded single photons with fair sampling is also shown in the figure for comparison. It can be seen that the results with coherent states are in strong contrast to those with single photons, where $E_1$, $E_2$ and $E_3$ give nearly the same violation of the inequality.
\par
The definition of the measurement event is essential for testing contextuality with classical systems. Actually, we assign conditional probabilities $P(A_i=a_i,A_{i+1}=a_{i+1}|E_j)$ under $E_j$. Since in general $E_1\subset E_2 \subset E_3$, $E_1$ and $E_2$ would register events \textit{conditionally}, resulting in post-selection in measurement outcomes. With such post-selection, the probabilities of events are observed and renormalized in a subspace of the complete outcomes. The post-selection probability under $E_j$ can be calculated by
\begin{equation}
P(E_j)=\frac{P(\{A_i=a_i,A_{i+1}=a_{i+1}\}\cap E_j)}{P(\{A_i=a_i,A_{i+1}=a_{i+1}\}|E_j)},\ \forall a_i,a_{i+1}.
\label{eq:post}
\end{equation}
Figure \ref{fig:results:b} demonstrates the post-selection probabilities $P(E_j)$ for different $E_j$~\cite{Supplementary}. $E_1$ always violates the inequality, but maintains a small post-selection probability. The value of $\beta$ under $E_2$ increases with the increase of the post-selection probability $P(E_2)$, while $P(E_2)$ approaches 1 at large $\bar{n}$, due to the vacuum component of the input field becomes negligible. These results demonstrate the quantum-to-classical transition in the contextuality test, from single photons to coherent light fields. For the full statistics $E_3$, which poses no post-selection, we observe no violation of the KCBS inequality.

\par
\textit{Discussion.}---The violation of the NC inequality is determined by the joint photon number distributions before the detectors in our setup. According to Eq.~(\ref{eq:correlation}), the violation is related to the correlation function of the outcomes $i,i+1$ defined by
\begin{equation}
g_{i,i+1}=\frac{P(A_i=-1,A_{i+1}=-1)}{P(A_i=-1)P(A_{i+1}=-1)}.
\label{eq:g}
\end{equation} 
This quantity is similar to the second order correlation function $g^{(2)}$ often used to assess the non-classicality of light field. For a linear-optical network with a coherent light field as the input, the output state is uncorrelated between different modes~\cite{PhysRevA.89.052302,PhysRevLett.117.213602}, \textit{i.e.} $g_{i,i+1}=1$. It follows that $\langle A_i A_{i+1}\rangle=[1-2P(A_i=-1)][1-2P(A_{i+1}=-1)]=\langle A_i \rangle\langle A_{i+1}\rangle$, and thus the KCBS inequality cannot be violated. Moreover, if the input field is not perfectly coherent (e.g., the thermal light), the output modes are classically correlated~\cite{PhysRevA.79.062303}, but the result still does not violate the NC inequality \cite{Supplementary}. Violating the KCBS inequality requires at least $g_{i,i+1}<1$, implying non-classical correlations of outcomes which put the requirement on input states. To elucidate this requirement, we start with the Glauber--Sudarshan $P$ representation of the input optical field $\hat{\rho}=\int d^2\alpha \  P(\alpha) |\alpha\rangle\langle\alpha|$~\cite{PhysRev.131.2766,PhysRevLett.10.277}. Then the value of $\beta$ can be represented as
\begin{equation}
\beta_{E_j}=\int  d^2\alpha \ P(\alpha) \beta \left(\alpha |E_j\right),
\label{eq:pfunction}
\end{equation}
where $\beta \left(\alpha |E_j\right)$ denotes the value of $\beta$ for the coherent state $|\alpha\rangle$ under $E_j$.  If one aims at testing the contextuality of the \textit{input} system, the measurement event should be correspondingly defined in accordance with the input ($E_3$) instead of the help of ancillary outcomes (see Fig. \ref{fig:framework}). For our setup $\beta \left(\alpha |E_3\right)=5 \left[1-2\exp\left(-|\alpha|^2/\sqrt{5}\right)\right]^2\geq 0$~\cite{Supplementary}, we have
\begin{equation}
\beta_{E_3} = 5 \int  d^2\alpha \ P(\alpha) \left[1-2\exp\left(-|\alpha|^2/\sqrt{5}\right)\right]^2,
\label{eq:pfunction_E3} 
\end{equation}
where $P(\alpha)$ uniquely determines the photon number distribution~\cite{Supplementary}. $\beta_{E_3}<-3$ requires $P(\alpha)$ to be negative for certain $\alpha$ or more singular than the Dirac-delta function. In this sense, violation of the KCBS inequality without post-selection provides a sufficient criteria for the non-classicality of the input optical state. As an example, we provide a numerical simulation on the violation of the KCBS inequality with different optical states (see Fig.~S4 of the Supplemental Material~\cite{Supplementary}).
\par
On the other hand, the measurement event definition procedure in our experiment can be understood as a post-selection on certain component of the input optical state. For example, $E_1$ effectively selects the single-photon component when the intensity is low, while $E_2$ removes the vacuum component of the state \cite{book:14435}. Such post-selection can be understood as the time reversal between input and measured states~\cite{PhysRevLett.98.223601}, and can be confirmed by comparing $\beta_{E_1}$ and $\beta_{E_2}$ of coherent states with $\beta_{E_3}$ of the post-selected components, respectively~\cite{Supplementary}. Both the post-selected states possess non-classical superposition in the $P$ function, allowing a possible violation of the KCBS inequality. Indeed the post-selected correlation $g_{i,i+1}$ for $E_1$ always equals 0, while $g_{i,i+1}=1-\exp\left(-|\alpha|^2\right) < 1$ for $E_2$. From this perspective, the violation of the inequality should be attributed to the post-selected non-classical components, which are non-deterministically generated with probability $P(E_j)<1$. In particular for the coherent state input, $P(E_1) = \exp\left(-2|\alpha|^2/\sqrt{5}\right)+2\exp\left[-\left(\sqrt{5}-1\right)|\alpha|^2/\sqrt{5}\right]-3\exp(-|\alpha|^2)$ and $P(E_2) = 1-\exp\left(-|\alpha|^2\right)$. Accordingly, the resource for this non-classical behavior should be carefully evaluated~\cite{PhysRevLett.107.113603}. One can demonstrate quantum (even beyond quantum) contextual statistics with only classical inputs but at a cost of more resources, \textit{i.e.} more copies of states, due to the reduced post-selection probability. The relation between this resource and the memory cost needed to demonstrate contextuality~\cite{1367-2630-13-11-113011,PhysRevLett.120.130401} is an interesting issue to be further investigated.
\par
\textit{Conclusion.}---We show how the definition of measurement events affects the emergence of the violation of the KCBS inequality for both single photons and coherent states with linear-optical setup. It is shown that the violation of the non-contextual inequality requires non-classicality in the quasi-distribution of the optical states in phase space. Quantum correlations can emerge by filtering the non-classical component of the optical field through an appropriate definition of measurement events, but at a cost of reduced post-selection probability. Our results shed new light on the boundary between the quantum and classical domains, and foreshadow a new means of resource evaluation for classical simulation of quantum contextuality.

\begin{acknowledgments}
We thank X.-F. Qian, J. Sperling and I. A. Walmsley for fruitful discussions. A. Zhang acknowledges Y. Xu for valuable comments on this manuscript. This work was supported by the National Key Research and Development Program of China under Grant No. 2017YFA0303703, the National Natural Science Foundation of China under Grants No. 11690032, No. 61490711, No. 11474159 and No. 11574145. M. S. Kim is grateful to the financial support from EPSRC (EP/K034480/1), the KIST Institutional Program (2E26680-18-P025), Samsung GRO grant and Royal Society.
\end{acknowledgments}

%
\clearpage
\begin{widetext}
\section{Supplemental Material}
\section{Experimental details}
\par
\textit{Single photon source and classical light source}.---Light pulses of $\sim$150 fs duration, centered at 830 nm, originating from a Ti:Sapphire laser (76MHz repetition rate, Coherent Mira-HP) are split by a beam splitter into two beams. The transmitted one is sent to a pulse picker (Coherent, distinction ratio $> 500:1$) since the repetition rate of the laser oscillator is too high to the bandwidth of photon-counting detectors. The repetition rate of the light pulses is reduced to 1.9MHz and the intensity of the pulses is attenuated by a set of neutral density filters. A synchronizing signal from the pulse picker (1.9MHz) gives the trigger signal of the weak coherent state input. The reflected beam is frequency doubled in a $\beta-$barium borate (BBO) crystal to generate second harmonic beam (415 nm central wavelength). Then the beam is used to pump a 10 mm bulk potassium dihydrogen phosphate (KDP) crystal phase-matched for type-II collinear degenerate downconversion, which produce photon pairs, denoted as signal and idler. The signal and idler photons are separated by a polarizing beam splitter (PBS) and coupled into single mode fiber. The idler mode is detected by an avalanche photodiode (Excelitas Technologies, SPCM-AQRH-FC with a detection efficiency about 49\%) to give the trigger of the heralded single photon source, while the signal mode is sent to the measurement device shown in the main text. In addition, the second order coherence value $g^{(2)}(0)$ of the heralded single photons is measured by splitting the signal mode at a 50:50 beam-splitter and detecting the two output modes with two APDs (labeled by $a,b$), then the second order coherence value can be calculated by
\begin{equation}
g^{(2)}(0)=\frac{C_{a,b,T}}{C_{a,T}C_{b,T}}N_T.
\end{equation}
Here $C_{a,b,T}$ denotes the three-fold coincidence rate between detectors $a,b$ and heralding, while $C_{a,T}$ ($C_{b,T}$) denotes the coincidence rate between detectors $a$ ($b$) and heralding. The experimental result exhibits $g^{(2)}(0)=0.0397(38)$, ensuring the measured state is very close to a single photon state. Moreover, to estimate the imperfections in the state preparation and detection, and thereby the vacuum component caused by loss in the experiment, we calibrate the losses in the state preparation and detection respectively, and reconstruct the photon number statistics of the state at three different positions of the setup \cite{Daryl2004,PhysRevLett.97.043602}: (i) before all the loss, i.e., derived from the final photon number distribution and the beam splitter model; (ii) before the measurement device, accounting for the imperfection in the source; (iii) accounting for all the loss. The results are shown in Fig. \ref{fig:sta}.
\begin{figure}[h]
\centering
\addtocounter{figure}{1}
\subfigure{\includegraphics[width=0.3\linewidth]{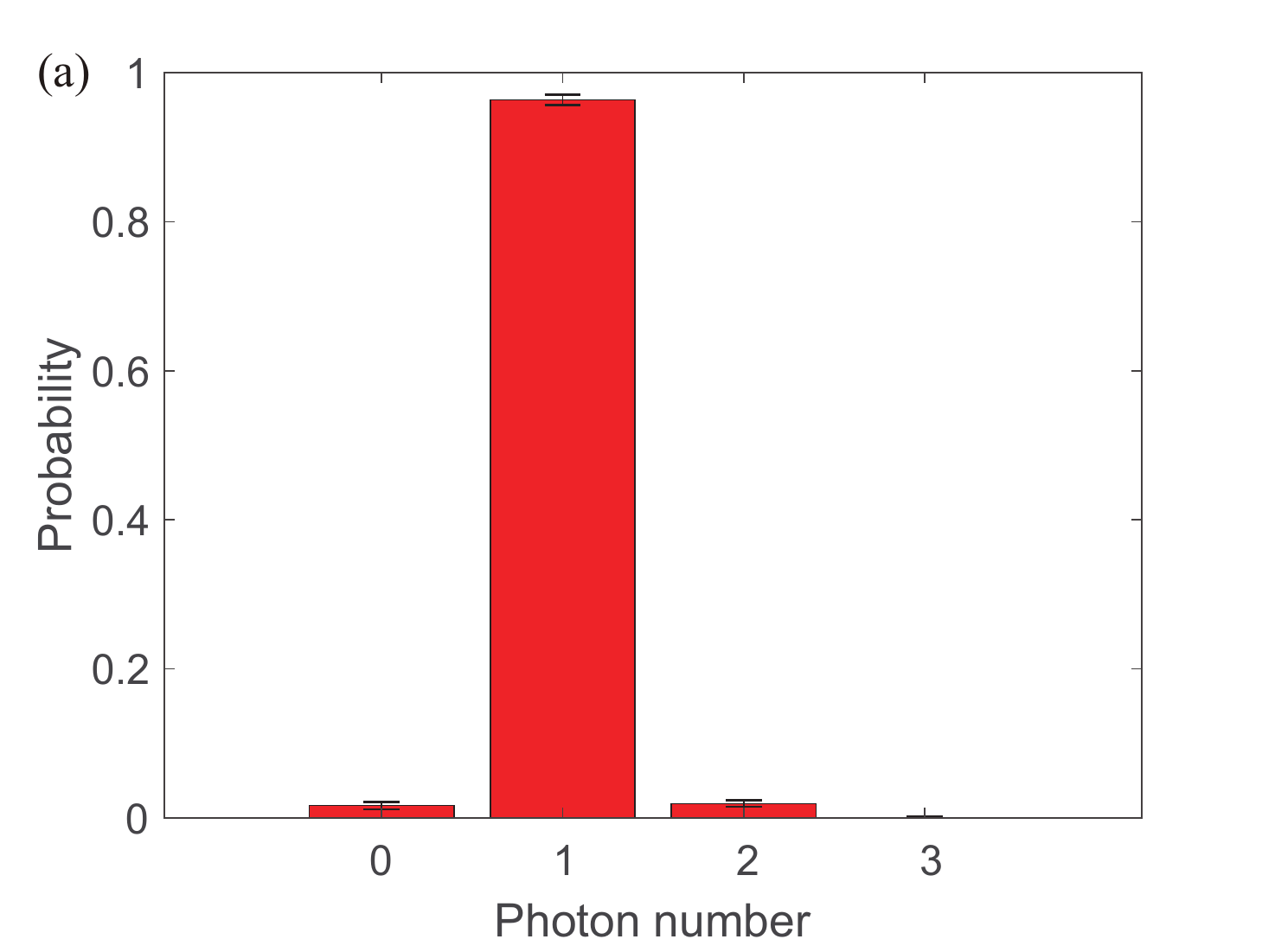}\label{fig:sta:a}}
\quad
\subfigure{\includegraphics[width=0.3\linewidth]{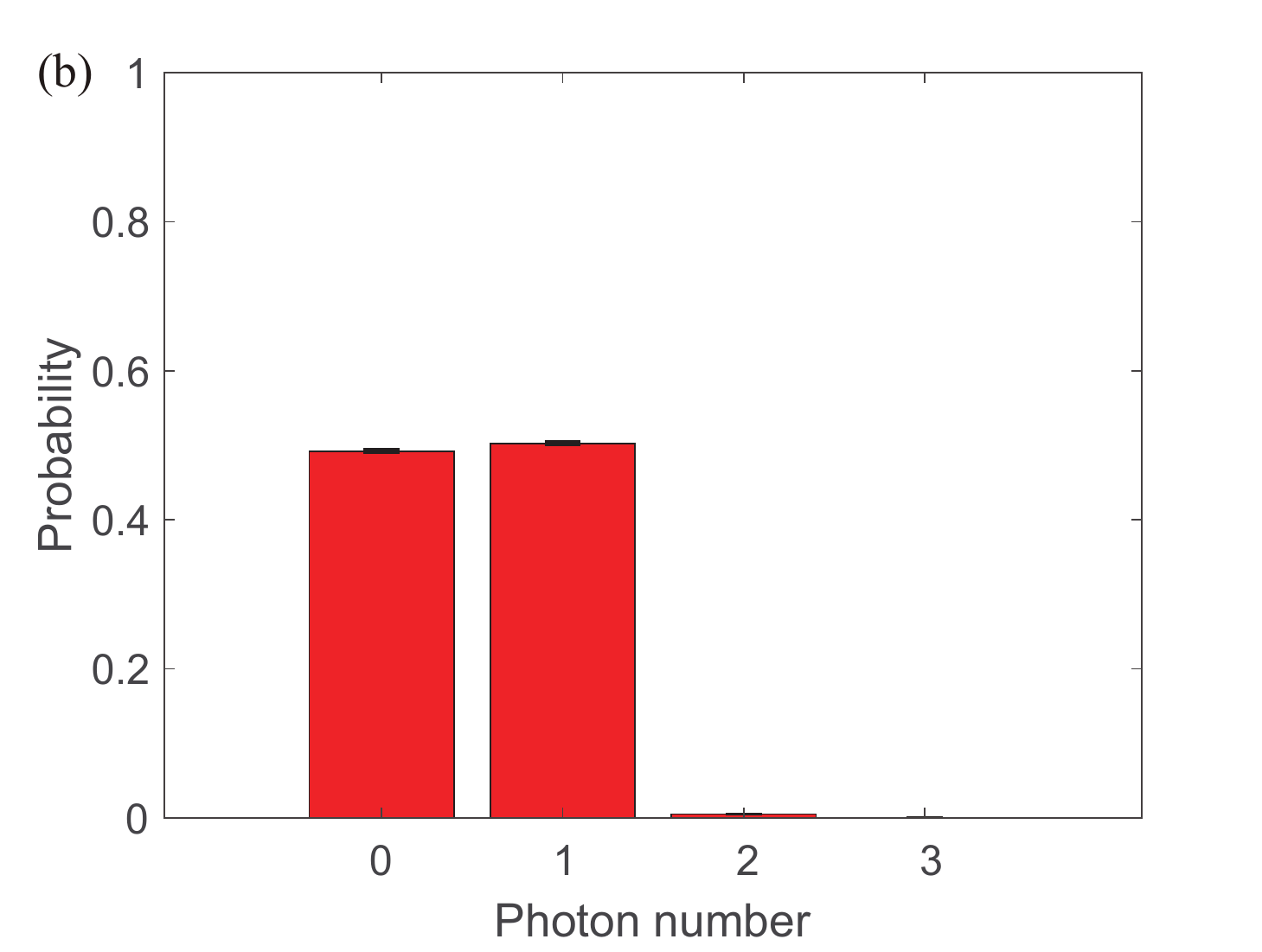}\label{fig:sta:b}}
\quad
\subfigure{\includegraphics[width=0.3\linewidth]{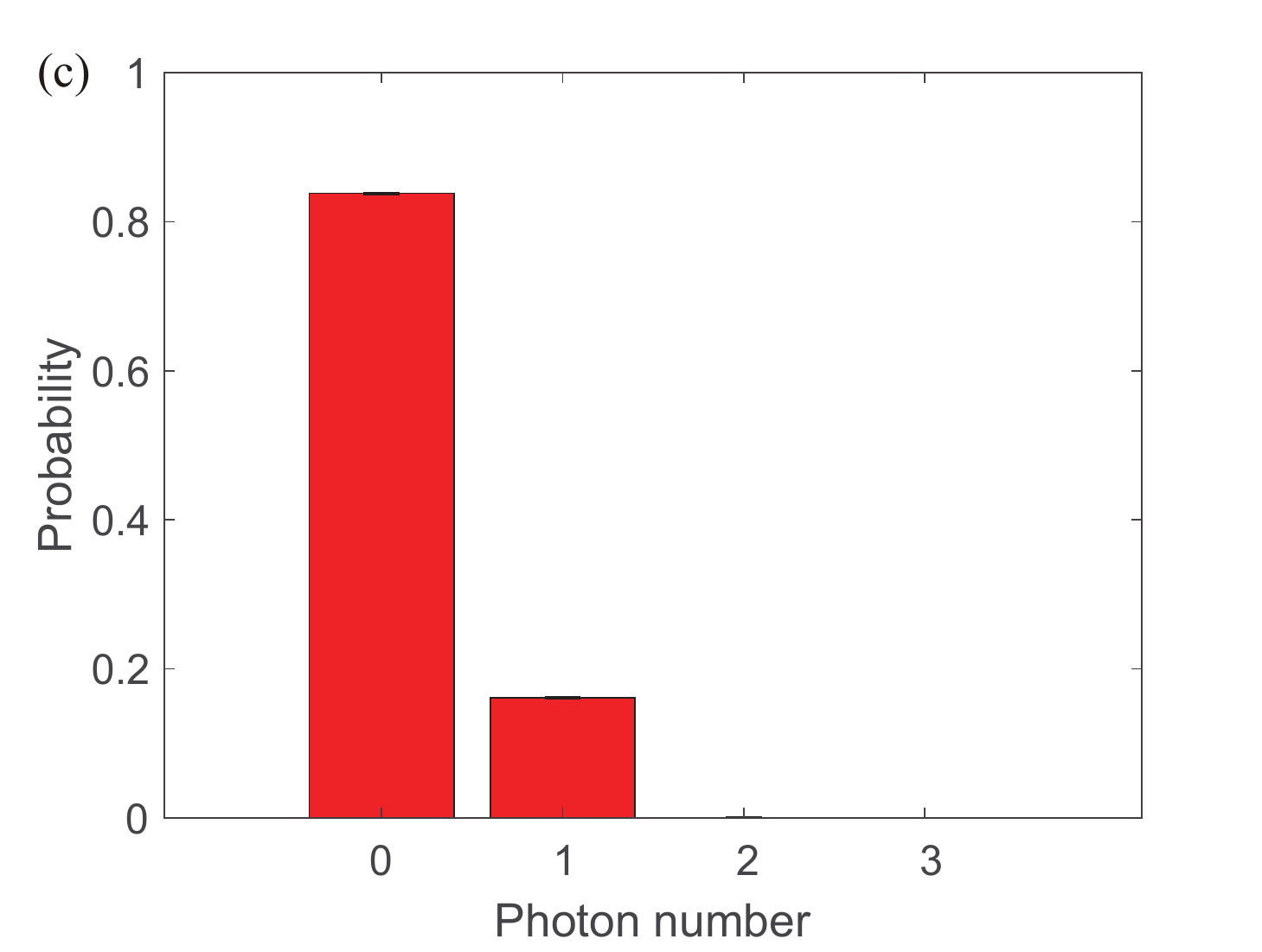}\label{fig:sta:c}}
\addtocounter{figure}{-1}
\caption{\label{fig:sta} The reconstructed photon statistics of the single photon source at different positions: (a) before all the loss,(b) before the measurement device and (c) accounting for all the loss.}
\end{figure}
\par
\textit{Fair sampling for the two sources}.---The idea of fair sampling is that the reported statistics is a fair statistical sample of the results of kind of states without any loss, based on the assumption that the loss is independent on the measurement settings. For heralded single photons, the loss in the setup would add additional vacuum component to the state, and all the events with no clicks on the three detectors are caused by loss. As a result, we can extrapolate the measurement results to what would be obtained with $100\%$ detection efficiency according to the fair sampling assumption. The events conditioned on ``at least one of the three detectors ($D_1,D_2,D_3$) clicks, and simultaneously the trigger $D_T$ clicks'' are fair samples of the ensemble of all heralded single photons without loss. Alternatively, one can also consider the results without applying fair sampling. In the following, we consider an experiment with non-unit heralding efficiency, and the results correspond to a state including the vacuum component caused by losses in the apparatus. The heralding efficiency $\eta_H$ is defined by
\begin{equation}
\eta_H=\frac{\sum_i C_{i,T}}{N_T},
\end{equation}
where $N_T$ denotes the count rate of heralding clicks (the trigger detector $D_T$), and $C_{i,T}$ denotes the coincidence count rate between the detector $i$ and the trigger detector $D_T$. The measured heralding efficiency of single photons in the experiment is 16.2\%. Figure \ref{fig:HSPS:a} shows the KCBS value for the single photon source without applying fair sampling, which gives the measured value $\beta=3.5550(30)$ under $E_3$ and $\beta=-3.9176(60)$ under $E_2$ in our case. The result of $E_1$ is very close to that of $E_2$ because the multi-photon component for the source is nearly zero. The detailed experimental values are shown in Table. \ref{tab:hsps_e}. It can be seen from the figure that the increasing of the value is mainly due to the finite heralding efficiency compared with the one for ideal single photons. If we do not apply the fair sampling assumption, the violation of the non-contextual bound cannot be observed with a heralding efficiency less than $89.65\%$. The single photon source can achieve quantum bound without applying fair sampling if the heralding efficiency is sufficiently high, while the coherent states with varied amplitudes never violate the inequality under full statistics $E_3$, irrespective of detection efficiency and intensity. 
\begin{table}
 \centering \caption{\label{tab:hsps_e}Summary of the results for the single photon source with or without fair sampling.}
\begin{ruledtabular}
\begin{tabular}{cccc}
    
   &$E_1$&$E_2$&$E_3$  \\
    \hline
    With fair sampling & $-3.9284(59)$ & $-3.9176(60)$ & $-3.9176(60)$ \\
    Without fair sampling & $-3.9284(59)$ & $-3.9176(60)$ & $3.5550(30)$ 
   \end{tabular}
  
\end{ruledtabular}
    \end{table}
\begin{figure}
\centering
\addtocounter{figure}{1}
\subfigure{\includegraphics[width=0.4\linewidth]{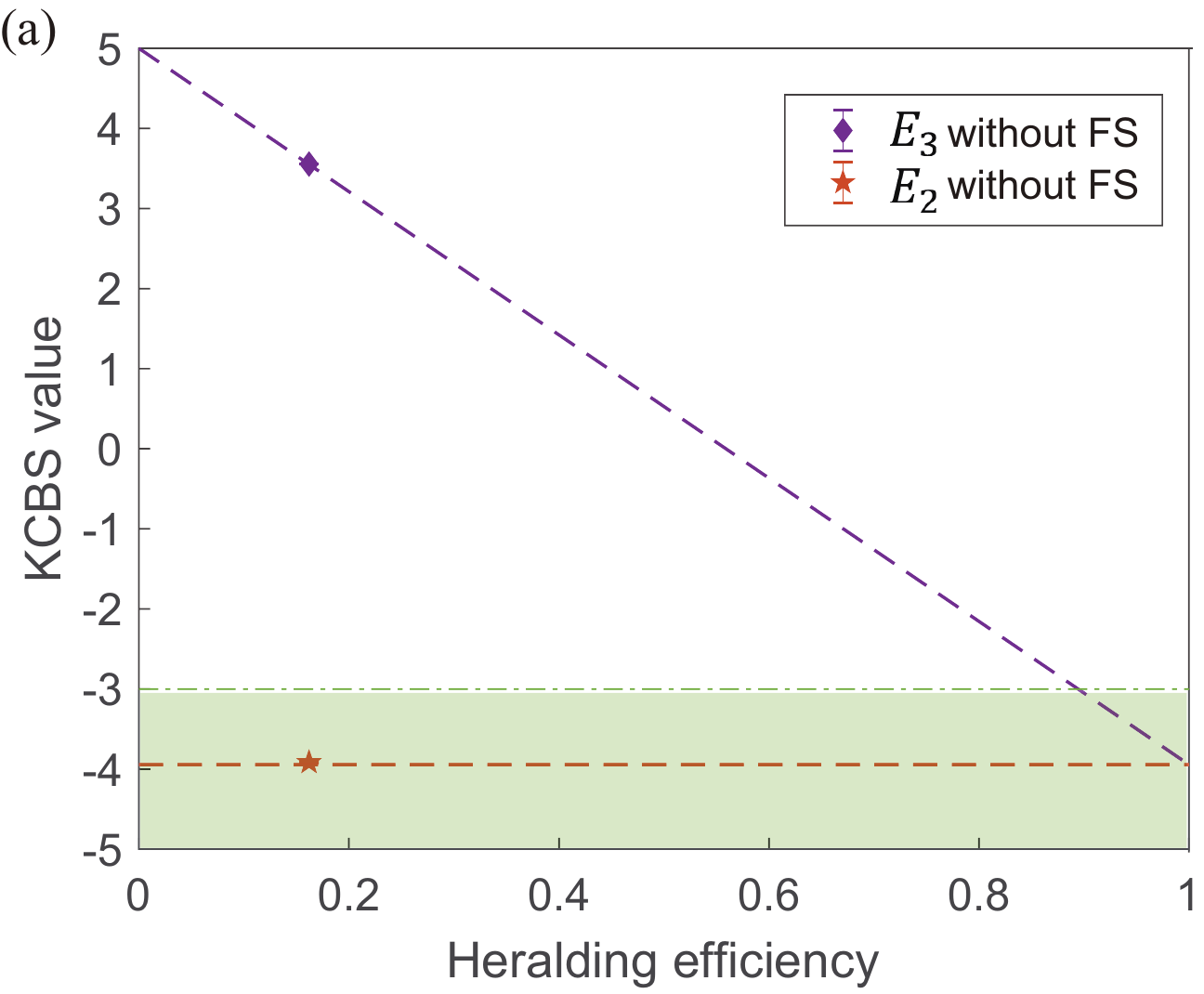}\label{fig:HSPS:a}}
\quad
\subfigure{\includegraphics[width=0.4\linewidth]{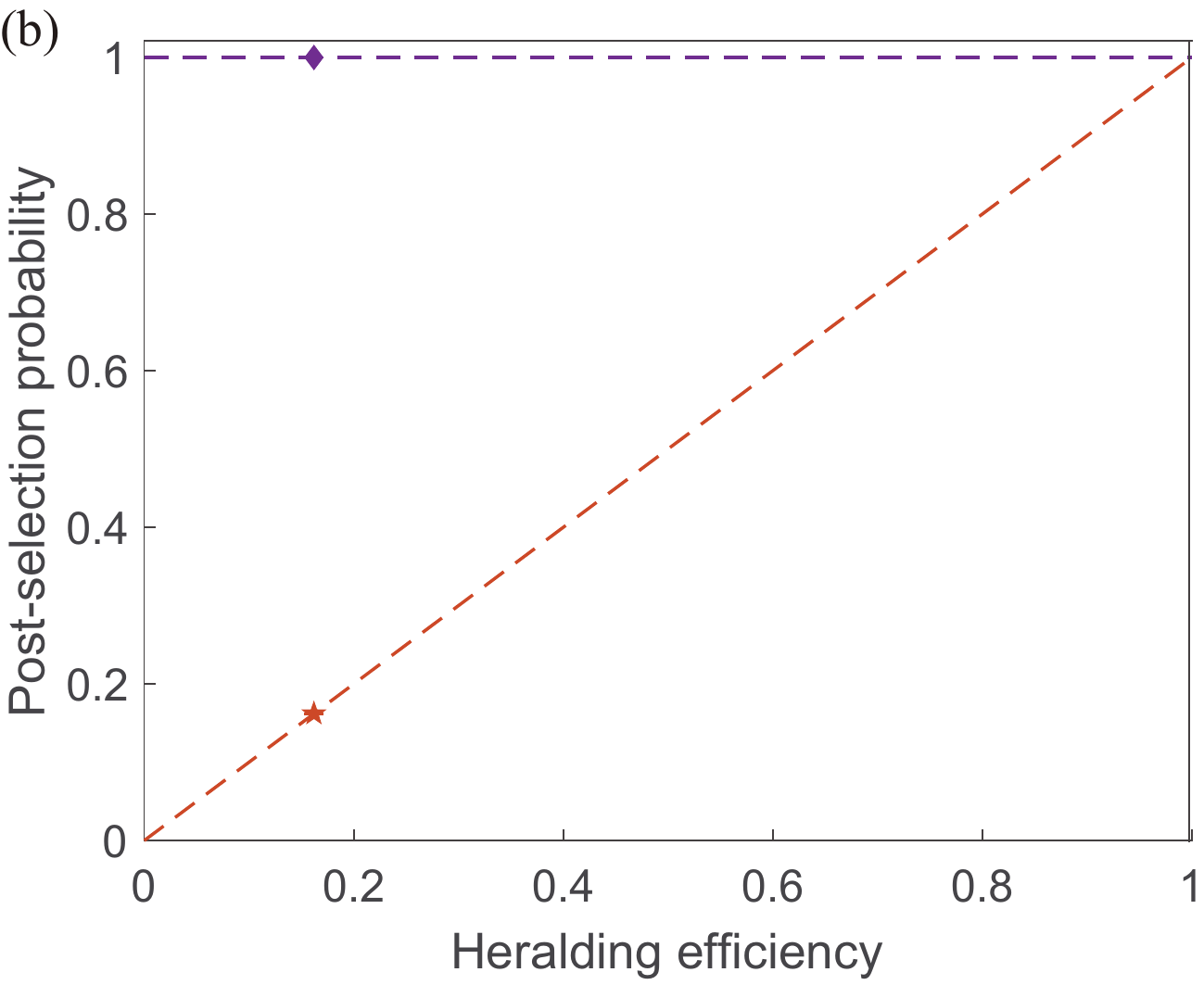}\label{fig:HSPS:b}}
\addtocounter{figure}{-1}
\caption{\label{fig:HSPS} The KCBS values and post-selection probabilities for heralded single photons without applying fair sampling: (a) the KCBS values, (b) The post-selection probabilities. }
\end{figure}
\begin{figure}[h]
  \centering
  \includegraphics[width=0.5\linewidth]{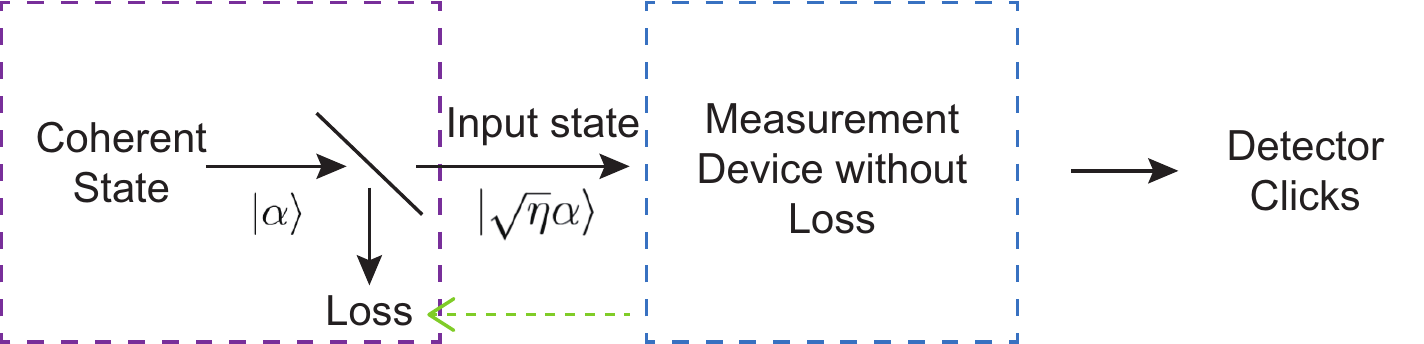}
\caption{\label{fig:splitter}The schematic diagram of fair sampling for coherent state inputs.}
\end{figure} 
\par
For coherent state input, fair sampling can be taken in the same idea that the loss is independent of measurement setting, thus the registered statistics is a representative sample of a attenuated coherent state accounting for the losses. Note a coherent state remains coherent with loss and the photon number distribution is always Poissonian with a smaller average photon number. Given that the measurement device is based on a linear optical network, it is reasonable to incorporate the overall losses from the photonic network and detection into the preparation part. The overall detection efficiency $\eta$ can be abstracted as a beam splitter with splitting ration $\eta:(1-\eta)$ in the preparation stage, thus the measurement result is equivalent to that for a coherent state $|\sqrt{\eta} \alpha\rangle$, and the linear optical network can be regarded as a lossless channel, as shown in Fig. \ref{fig:splitter}. The registered detector clicks is a fair sample of the coherent state with a smaller average photon number.
\par
To estimate the reduced average photon number of the coherent state $\eta|\alpha|^2$ from the click statistics, we consider the detection of coherent states $|\alpha_i\rangle$ distributed in $m$ modes. The average collected clicks when we put $m$ APDs to detect the light fields can be derived from the Poissonian distribution of coherent states
\begin{equation}
n_{\text{collected}}=\sum_{i=1}^m n'_i=\sum_{i=1}^m\left[1-\exp{\left(-|\alpha_i|^2\right)}\right].
\label{eq:number}
\end{equation}
Here $|\alpha_i|^2$ denotes the average photon number of the coherent state in the $i$-th mode, and $n'_i$ denotes the corresponding collected average clicks by the $i$-th APD. Applying Eq. (\ref{eq:number}) to the scenario in our experiment, we arrive at the average photon number $\bar{n}$ of the coherent state $|\sqrt{\eta} \alpha\rangle$
\begin{equation}
\bar{n}=\eta |\alpha|^2=\sum_{i=1}^m|\alpha_i|^2=\sum_{i=1}^m\ln{\frac{1}{1-n'_i}}.
\end{equation}

\begin{table}
 \centering \caption{\label{tab:setting}The correspondence between the registered events and the clicks of detectors $D_1,D_2$ and $D_3$ (as in Fig. 2 of the main text). Note all of the events are conditioned on the measurement event $E_j$.}
\begin{ruledtabular}
\begin{tabular}{cccccc}
   $i$  & \textbf{Measurement context} & $N(A_i=-1)$ & $N(A_{i+1}=-1)$ & $N(A_i=-1,A_{i+1}=-1)$  \\
    \hline
    1 & $\{A_1,A_2\}$ & $D_2$ & $D_3$& $D_2\cap D_3$ \\
    2 & $\{A_2,A_3\}$ & $D_3$ & $D_1$& $D_1\cap D_3$ \\
    3 & $\{A_3,A_4\}$ & $D_1$ & $D_2$& $D_1\cap D_2$ \\
    4 & $\{A_4,A_5\}$ & $D_2$ & $D_3$& $D_2\cap D_3$ \\
    5 & $\{A_5,A_1\}$ & $D_3$ & $D_2$& $D_2\cap D_3$ \\
   \end{tabular}
  
\end{ruledtabular}
    \end{table}
\par
\textit{Measurement setting and data collection}.---In the language of quantum theory, the maximum violation of the KCBS inequality can be achieved by performing 5 projective measurements on a qutrit system. For the single-photon input $|\psi\rangle=(0,0,1)^T$, the 5 projectors can be represented as $P_i=|\psi_i\rangle\langle\psi_i|$ ($i=1,2,...,5$), where
\begin{equation}
|\psi_i\rangle=c_i\left(\cos(4\pi i/5),\sin(4\pi i/5),\sqrt{\cos(\pi/5)}\right)^T.
\end{equation}
Here $c_i$ is the normalization coefficient. Then the 5 observables can be defined by $A_i=2P_i-\openone$, producing outcomes $\{+1,-1\}$ \cite{Klyachko_2008}. The three modes in our setup are defined as one spatial mode of the light field, and the vertical polarization and horizontal polarization in another spatial mode, denoted as mode 0,1,2 respectively. The beam displacers (BDs) and half wave-plates (HWPs) are then used to perform operations on the polarization modes and transformations on the spatial modes. The following detectors thus perform the corresponding measurements by the detection of the light field distributed in the three modes. The experiment goes by changing one of the measurement to another and measuring a different context. The data of the experiment is recorded by an event register, including the clicks of the detectors (the three APDs) and the signal from the input trigger. The trigger signal of the single photon source is the clicks of the trigger APD in the idler mode of the spontaneous parametric down-conversion (SPDC), while the trigger signal of the classical light pulses is a synchronizing signal of the pulse picker monitored by a fast photodiode (PD). In each measurement context, two of the three APDs give the outcomes of the observables $A_i, A_{i+1}$, while the third APD plays the role of ancillary outcomes $A_\text{anc}=0,1$. The detailed settings that the clicks of which two detectors are registered are summarized in Table. \ref{tab:setting}. For example, in the context $\{A_1,A_2\}$, we register the clicks of Detector 2 for the counts $n(A_1=-1)$, the clicks of Detector 3 for the counts $n(A_2=-1)$, and the coincidences of Detector 2 and Detector 3 for counts $n(A_i=-1,A_{i+1}=-1)$. All of the counts are collected with respect to the measurement event $E_j$ with counts $N_j$. Then the conditional probabilities $P(A_i=-1,A_{i+1}=-1|E_j)$ can be calculated by $P(A_i=-1,A_{i+1}=-1|E_j)=n(A_i=-1,A_{i+1}=-1|E_j)/N_j$. In other words, the probabilities are assigned and renormalized over the event $E_j$. In the experiment with full statistics ($E_3$), the statistics of outcomes are registered with respect to the input trigger without the help of the ancillary outcome. The dark noise is negligible compared with the count rate because all the registered counts are based on the coincidences between the detectors and the input trigger. 

\par
\begin{table}
 \centering \caption{\label{tab:result_sp}Collected experimental result for the single photon source input.}
\begin{ruledtabular}
\begin{tabular}{ccccccc}
   $i$  & \textbf{Measurement context} & $P(A_i=-1)$ & $P(A_{i+1}=-1)$ & $P(A_i=-1,A_{i+1}=-1)$  &&  \textbf{Calculated value} $\langle A_i A_{i+1}\rangle$\\
    \hline
   1 & $\{A_1,A_2\}$ & $0.4446(19)$ &0.4495(19) & $0.0013(2)$ && $-0.7831(28)$\\
   2 & $\{A_2,A_3\}$ & $0.4593(20)$ &0.4442(21) & $0.0013(2)$  && $-0.8017(27)$\\
   3 & $\{A_3,A_4\}$ & $0.4570(20)$ &0.4323(19)& $0.0013(1)$  && $-0.7735(24)$\\
   4 & $\{A_4,A_5\}$ & $0.4290(22)$ &0.4583(22) & $0.0013(2)$  && $-0.7692(28)$\\
   5 & $\{A_5,A_1\}$ & $0.4573(20)$ & 0.4403(20) & $0.0013(1)$  && $-0.7900(28)$\\
   \hline
   &&&&&$\beta$&$-3.9176(60)$
  \end{tabular}
\end{ruledtabular}
    \end{table}
\textit{Summary of results and imperfections}.---As a detailed summary of the experimental results, the experimental values with error bars for the coherent state inputs are summarized in Table. \ref{tab:results}. Moreover, we give the collected experimental probabilities for several representative cases with coherent states under different measurement events $E_j$, as shown in Tables. \ref{tab:result_01} and \ref{tab:result_124}. For each measurement context, the assigned probabilities $P(A_i=-1|E_j)$, $P(A_{i+1}=-1|E_j)$ and $P(A_i=-1,A_{i+1}=-1|E_j)$ with respect to the measurement event $E_j$ are used to calculate the correlation $\langle A_i A_{i+1}\rangle_{|E_j}$ by Eq. (2) in the main text.
 \begin{table}[h]
 \centering \caption{\label{tab:bound}Collected experimental result to correct the bound.}
\begin{ruledtabular}    
   \begin{tabular}{ccccc}    
    $P(A_1=-1)$ & $P(A'_1=1|A_1=-1)$ & $P(A_1=1)$ & $P(A'_1=-1|A_1=1)$ &\textbf{Calculated} $-3-\epsilon$\\
    \hline
     $0.4446(19)$ & $0.0058(7)$ &0.5554(19)& $0.0020(4)$  &  $-3.0074(9)$\\
   \end{tabular}
   \end{ruledtabular}
    \end{table}
\par
In our experiment, the main imperfection is the inaccuracies of the configurations of waveplates and beam displacers. This imperfection may lead to five observables which are not ideally symmetric as the maximum violation case. As a result, the measured probabilities do have some discrepancies from the ideal case, resulting in a lower observed violation of the KCBS inequality. However, this imperfection does not change the non-contextual bound [Eq. (1) in the main text], because the derivation of the non-contextual bound does not depend on any specific arrangement of the experiment, only the axioms of probability theory are assumed. An observed violation of the non-contextual inequality always shows the contextuality of the system, therefore our arguments are not affected.
\par
Another issue is that due to the experimental imperfections, the measurement $A_1$ in the context $\{A_1,A_2\}$ (denoted as $A_1$ in the following) and context $\{A_5,A_1\}$ (denoted as $A_1'$ in the following) are not physically the same. To take this imperfection into consideration, we adopted the method proposed in \cite{Lapkiewicz_2011} to characterize the difference between $A_1$ and $A_1'$ and correct the non-contextual bound. The non-contextual inequality in the practical case is a test including six observables, $A_1, A_2, ..., A_5$ and $A_1'$. The effect of this difference on the non-contextual bound is quantified by
\begin{equation}
\epsilon=1-\langle A_1 A_1' \rangle=2\left[P(A_1'=-1|A_1=1)P(A_1=1)+P(A_1'=1|A_1=-1)P(A_1=-1)\right].
\end{equation}
The conditional probabilities $P(A_1'=a_1'|A_1=a_1)$ can be experimentally measured by setting the measurement context in $\{A_5,A_1'\}$ and blocking the mode corresponding to $A_1$ or the other modes except for $A_1$. As a result, the experiment is conducted under the condition $A_1=1$ or $A_1=-1$. The experimental result to correct the bound is shown in Table. \ref{tab:bound}. This difference render the actual non-contextual slightly different from the ideal case (-3). All the results in the experiment are compared with the characterized bound rather than the theoretical one.

\begin{table}
 \centering \caption{\label{tab:results}Experimental results for classical light field (coherent state) inputs under different definitions of measurement events. $\beta_\text{th}$, $\beta_\text{exp}$ represent the theoretical and experimental values of the left hand of the KCBS inequality respectively, and $P_\text{th}(E_j)$, $P_\text{exp}(E_j)$ represent the theoretical and experimental values of post-selection probabilities respectively.}
\begin{ruledtabular}
\begin{tabular}{ccccccc}
    &\multicolumn{2}{c}{\textbf{$E_1$: Only 1 detector clicks}}&\multicolumn{2}{c}{\textbf{$E_2$: At least 1 detector clicks}}&\multicolumn{2}{c}{\textbf{$E_3$: Full statistics}}\\
   $\bar{n}$  & $\beta_\text{th}$ & $\beta_\text{exp}$ & $\beta_\text{th}$ & $\beta_\text{exp}$ & $\beta_\text{th}$ & $\beta_\text{exp}$ \\
    \hline
0.10 & -3.9611 & -3.9471(32) & -3.7830 & -3.7680(32) & 4.1296 & 4.1315(24) \\
0.40 & -4.0078 & -3.9788(39) & -3.3129 & -3.2858(31) & 2.2720 & 2.2840(41) \\
0.72 & -4.0579 & -4.0146(22) & -2.7798 & -2.7448(21) & 1.0192 & 1.0462(37) \\
0.99 & -4.1003 & -4.0574(28) & -2.3081 & -2.2716(40) & 0.4009 & 0.4404(60) \\
1.24 & -4.1378 & -4.1018(34) & -1.8802 & -1.8478(38) & 0.1103 & 0.1488(21) \\
1.57 & -4.1859 & -4.1482(15) & -1.3215 & -1.2843(71) & 0.0002 & 0.0506(14) \\
1.84 & -4.2260 & -4.1843(21) & -0.8503 & -0.8069(73) & 0.0756 & 0.1352(28)\\
    \hline
    &&&&&&\\
     
   $\bar{n}$  & $P_\text{th}(E_1)$ & $P_\text{exp}(E_1)$ & $P_\text{th}(E_2)$ & $P_\text{exp}(E_2)$ & $P_\text{th}(E_3)$ & $P_\text{exp}(E_3)$\\
    \hline
0.10 & 0.0961 & 0.0960(3) & 0.0991 & 0.0991(3)  & 1 & 1 \\
0.40 & 0.2904 & 0.2896(4) & 0.3282 & 0.3278(5)  & 1 & 1 \\
0.72 & 0.4075 & 0.4049(3) & 0.5117 & 0.5105(5)  & 1 & 1 \\
0.99 & 0.4551 & 0.4505(4) & 0.6293 & 0.6271(11) & 1 & 1 \\
1.24 & 0.4694 & 0.4647(2) & 0.7107 & 0.7084(6)  & 1 & 1 \\
1.57 & 0.4614 & 0.4552(3) & 0.7909 & 0.7877(8)  & 1 & 1 \\
1.84 & 0.4394 & 0.4316(4) & 0.8417 & 0.8378(6)  & 1 & 1
   \end{tabular}
\end{ruledtabular}
    \end{table}
 
    \begin{table}
 \centering \caption{\label{tab:result_01}Collected experimental result for classical light input with $\bar{n}=0.10$ under different measurement events $E_j$.}
\begin{ruledtabular}
\begin{tabular}{ccccccc}
   $i$  & \textbf{Measurement context} & $P(A_i=-1|E_1)$ & $P(A_{i+1}=-1|E_1)$ & $P(A_i=-1,A_{i+1}=-1|E_1)$  &&  \textbf{Calculated value} $\langle A_i A_{i+1}\rangle_{|E_1}$\\
    \hline
1 & $\{A_1,A_2\}$ & 0.4455(10) & 0.4468(10) & 0(0) && -0.7845(11) \\
2 & $\{A_2,A_3\}$ & 0.4502(10) & 0.4525(10) & 0(0) && -0.8054(12) \\
3 & $\{A_3,A_4\}$ & 0.4576(11) & 0.4334(10) & 0(0) && -0.7820(13) \\
4 & $\{A_4,A_5\}$ & 0.4291(12) & 0.4583(14) & 0(0) && -0.7748(17) \\
5 & $\{A_5,A_1\}$ & 0.4591(10) & 0.4411(10) & 0(0) && -0.8004(13) \\
    \hline
    &&&&&$\beta$&$-3.9471(32)$\\
   &&&&&&\\
   $i$  & \textbf{Measurement context} & $P(A_i=-1|E_2)$ & $P(A_{i+1}=-1|E_2)$ & $P(A_i=-1,A_{i+1}=-1|E_2)$  &&  \textbf{Calculated value} $\langle A_i A_{i+1}\rangle_{|E_2}$\\
    \hline
1 & $\{A_1,A_2\}$ &0.4578(9)  & 0.4591(10) & 0.0213(3) && -0.7487(12) \\
2 & $\{A_2,A_3\}$ &0.4626(10) & 0.4649(11) & 0.0218(3) && -0.7678(13) \\
3 & $\{A_3,A_4\}$ &0.4696(11) & 0.4459(9)  & 0.0211(3) && -0.7466(13) \\
4 & $\{A_4,A_5\}$ &0.4414(12) & 0.4701(13) & 0.0206(3) && -0.7405(18) \\
5 & $\{A_5,A_1\}$ &0.4710(10) & 0.4533(10) & 0.0211(3) && -0.7643(13)\\
    \hline
    &&&&&$\beta$&$-3.7680(32)$\\
    &&&&&&\\
   $i$  & \textbf{Measurement context} & $P(A_i=-1|E_3)$ & $P(A_{i+1}=-1|E_3)$ & $P(A_i=-1,A_{i+1}=-1|E_3)$  &&  \textbf{Calculated value} $\langle A_i A_{i+1}\rangle_{|E_3}$\\
    \hline
    1 & $\{A_1,A_2\}$ & $0.0459(3)$ &0.0460(3) & $0.0021(0)$ && $0.8248(9)$\\
    2 & $\{A_2,A_3\}$ & $0.0464(4)$ &0.0466(4) & $0.0022(0)$  && $0.8227(14)$\\
    3 & $\{A_3,A_4\}$ & $0.0465(5)$ &0.0442(4)& $0.0021(0)$  && $0.8269(17)$\\
    4 & $\{A_4,A_5\}$ & $0.0434(5)$ &0.0462(4) & $0.0020(0)$  && $0.8290(16)$\\
    5 & $\{A_5,A_1\}$ & $0.0459(4)$ & 0.0442(4) & $0.0021(0)$  && $0.8280(13)$\\
    \hline
    &&&&&$\beta$&$4.1315(24)$\\
   \end{tabular}
\end{ruledtabular}
    \end{table}
    
 \begin{table}
 \centering \caption{\label{tab:result_124}Collected experimental result for classical light input with $\bar{n}=1.24$ under different measurement events $E_j$.}
\begin{ruledtabular}
\begin{tabular}{ccccccc}
   $i$  & \textbf{Measurement context} & $P(A_i=-1|E_1)$ & $P(A_{i+1}=-1|E_1)$ & $P(A_i=-1,A_{i+1}=-1|E_1)$  &&  \textbf{Calculated value} $\langle A_i A_{i+1}\rangle_{|E_1}$\\
    \hline
1 & $\{A_1,A_2\}$ & 0.4435(5) & 0.4677(5)  & 0(0) && -0.8223(5)  \\
2 & $\{A_2,A_3\}$ &0.4674(6) & 0.4489(6)  & 0(0) && -0.8327(6)  \\
3 & $\{A_3,A_4\}$ &0.4634(5) & 0.4434(5)  & 0(0) && -0.8137(7)  \\
4 & $\{A_4,A_5\}$ &0.4395(7) & 0.4679(6)  & 0(0) && -0.8148(8)  \\
5 & $\{A_5,A_1\}$ &0.4637(10) & 0.4454(15) & 0(0) && -0.8183(30)\\
    \hline
    &&&&&$\beta$&$-4.0574(28)$\\
   &&&&&&\\
   $i$  & \textbf{Measurement context} & $P(A_i=-1|E_2)$ & $P(A_{i+1}=-1|E_2)$ & $P(A_i=-1,A_{i+1}=-1|E_2)$  &&  \textbf{Calculated value} $\langle A_i A_{i+1}\rangle_{|E_2}$\\
    \hline
1 & $\{A_1,A_2\}$ &0.5899(9)  & 0.6080(10) & 0.2563(16) && -0.3708(29) \\
2 & $\{A_2,A_3\}$ &0.6104(8) & 0.5967(13) & 0.2618(17) && -0.3668(28) \\
3 & $\{A_3,A_4\}$ &0.6040(4) & 0.5888(5)  & 0.2539(5) && -0.3699(9) \\
4 & $\{A_4,A_5\}$ &0.5851(16) & 0.6064(9) & 0.2524(21) && -0.3734(35) \\
5 & $\{A_5,A_1\}$ &0.6057(8) & 0.5920(16) & 0.2571(12) && -0.3669(13)\\
    \hline
    &&&&&$\beta$&$-2.2716(40)$\\
    &&&&&&\\
   $i$  & \textbf{Measurement context} & $P(A_i=-1|E_3)$ & $P(A_{i+1}=-1|E_3)$ & $P(A_i=-1,A_{i+1}=-1|E_3)$  &&  \textbf{Calculated value} $\langle A_i A_{i+1}\rangle_{|E_3}$\\
    \hline
    1 & $\{A_1,A_2\}$ & $0.4175(20)$ &0.4304(21) & $0.1814(17)$ && $0.0297(13)$\\
    2 & $\{A_2,A_3\}$ & $0.4348(21)$ &0.4251(24) & $0.1865(18)$  && $0.0263(17)$\\
    3 & $\{A_3,A_4\}$ & $0.4270(6)$ &0.4163(6)& $0.1795(5)$  && $0.0315(7)$\\
    4 & $\{A_4,A_5\}$ & $0.4121(29)$ &0.4271(24) & $0.1778(22)$  && $0.0326(18)$\\
    5 & $\{A_5,A_1\}$ & $0.4303(10)$ & 0.4207(16) & $0.1827(10)$  && $0.0288(9)$\\
    \hline
    &&&&&$\beta$&$0.4404(60)$\\
   \end{tabular}
\end{ruledtabular}
    \end{table}

\section{Details of theoretical predictions}
\par
According to the probability theory, the result of a correlation experiment can be abstracted as a mathematical description of the measurement scenario \cite{1367-2630-13-11-113036,PhysRevLett.119.050504}: the probability space. The probability space is described by a triple $(\Omega,E,P)$ which consists of 3 parts: the sample space $\Omega$, which is the set of all possible outcomes $\{a_i, a_{i+1}, a_\text{anc}\}$; the event space $E$, which is a set of events $\{e_k\}$ (subset of $\Omega$) relevant to the purpose of the experiment; the probabilities of events $P$, $P(\Omega)=1$. To test an non-contextual (NC) inequality, one can define a proper event space $E$, and then calculate the parameter $\beta$ with the collected probabilities $P$ of these events.
\par
In the following, we give the quantum description of the classical light experiment and derive the theoretical predictions. From the viewpoint of quantum theory, coherent states are the quantum-mechanical equivalents of classical monochromatic electromagnetic waves, due to the fact that the properties of coherent states most closely resemble the classical behavior of light field. Coherent states can be expanded in terms of the Fock states $|n\rangle$, and the probability of detecting $n$ photons subjects to the Poissonian distribution
\begin{eqnarray}
|\alpha\rangle=&&\sum_{n}\exp(-|\alpha|^2/2)\frac{\alpha^n}{(n!)^{1/2}}|n\rangle \\
\mathcal{P}(n)=&&|\langle n|\alpha\rangle|^2=\exp(-|\alpha|^2)\frac{|\alpha|^{2n}}{n!}.
\end{eqnarray}

\par
In the experiment with classical light fields, the input state can be denoted as
\begin{equation}
|\psi\rangle=|0\rangle_{0}|0\rangle_{1}|\alpha\rangle_{2},
\end{equation} 
where $0,1,2$ denote the 3 input optical modes. After the transformations of the linear optical network, the state becomes coherent states distributed in 3 modes. For example, in the measurement context $\{A_i,A_{i+1}\}$, the output state can be ideally described as
\begin{equation}
|\psi'\rangle=\left|\frac{1}{\sqrt[4]{5}}\alpha\right\rangle_{A_i}\left|\frac{1}{\sqrt[4]{5}}\alpha\right\rangle_{A_{i+1}}\left|\left(1-\frac{2}{\sqrt{5}}\right)^{1/2}\alpha\right\rangle_{A_\text{anc}}.
\end{equation} 
Here $A_\text{anc}$ denotes the optical mode of the ancillary observable corresponding to the third APD in the measurement device. We define $A_\text{anc}=0$ when the third APD gives a click and  $A_\text{anc}=1$ when it does not, then the measurement outcomes of the experiment can be represented of the form $\{a_i\ a_{i+1}\ a_\text{anc}\}$. Accordingly, the 3 conditions of measurement events $E_j$ can be defined by
\begin{eqnarray}
E_1=&&\{-1+1\ 1,+1-1\ 1,+1+1\ 0\}, \\
E_2=&&\{-1-1\ 1,-1-1\ 0,-1+1\ 1,-1+1\ 0,+1-1\ 1,+1-1\ 0,+1+1\ 0\}, \\
E_3=&&\Omega=\{-1-1\ 1,-1-1\ 0,-1+1\ 1,-1+1\ 0,+1-1\ 1,+1-1\ 0,+1+1\ 0,+1+1\ 1\}.
\end{eqnarray}
From the perspective of quantum measurements, the APDs perform projective measurements on the quantum state $|\psi'\rangle$. The projection operator $\hat{P}_{+1}=|0\rangle\langle 0|$ and $\hat{P}_{-1}=\openone-|0\rangle\langle 0|$, where $\openone$ is the identity operator. As a result, we can derive the joint probability distribution $P(A_i=a_i,A_{i+1}=a_{i+1}|E_j)$ over the outcomes under measurement context $\{A_i,A_{i+1}\}$ and measurement event $E_j$. Consequently, we can calculate the KCBS value $\beta (\alpha |E_j)=\sum_i M_i\left(\alpha |E_j\right)$ and the post-selection probability $P(E_j)$ for coherent state inputs. Here $i$ denotes different contexts $\{A_i,A_{i+1}\}$ and $M_i\left(\alpha |E_j\right)$ represents the measurement result $\langle A_iA_{i+1}\rangle_{|E_j}$ for the coherent state $|\alpha\rangle$ under $E_j$.
\par
For Condition (i):
\begin{eqnarray}
\beta (\alpha|E_1)=&&\sum_i M_i\left(\alpha |E_1\right)=5\times\frac{\exp\left(-2|\alpha|^2/\sqrt{5}\right)-2\exp\left[-\left(\sqrt{5}-1\right)|\alpha|^2/\sqrt{5}\right]+\exp(-|\alpha|^2)} {\exp\left(-2|\alpha|^2/\sqrt{5}\right)+2\exp\left[-\left(\sqrt{5}-1\right)|\alpha|^2/\sqrt{5}\right]-3\exp(-|\alpha|^2)}\\
P(E_1)=&&\exp\left(-2|\alpha|^2/\sqrt{5}\right)+2\exp\left[-\left(\sqrt{5}-1\right)|\alpha|^2/\sqrt{5}\right]-3\exp(-|\alpha|^2);
\end{eqnarray}
\par
For Condition (ii):
\begin{eqnarray}
\beta(\alpha|E_2)=&&\sum_i M_i\left(\alpha |E_2\right)=5\times\frac{1-\exp\left(-|\alpha|^2\right)+4\exp\left(-2|\alpha|^2/\sqrt{5}\right)-4\exp\left(-|\alpha|^2/\sqrt{5}\right)} {1-\exp\left(-|\alpha|^2\right)} \\
P(E_2)=&&1-\exp\left(-|\alpha|^2\right);
\end{eqnarray}
\par
For Condition (iii):
\begin{eqnarray}
\beta(\alpha|E_3)=&&\sum_i M_i\left(\alpha |E_3\right)=5\times\left[1-2\exp\left(-|\alpha|^2/\sqrt{5}\right)\right]^2 \\
P(E_3)=&&1.
\end{eqnarray}

\section{The correlation of the outcomes and the non-classicality of the state}
\begin{table}
 \centering \caption{\label{tab:g}The measured correlation function $g_{i,i+1}$ for the outcomes of $A_i,A_{i+1}$ for single photons and classical light fields.}
\begin{ruledtabular}
\begin{tabular}{cccccc}
    
     & $A_1 A_2$ & $A_2 A_3$ & $A_3 A_4$ & $A_4 A_5$ & $A_5 A_1$ \\
    \hline
    HSPS & $0.0391(48)$ & $0.0400(46)$ & $0.0413(44)$ & $0.0408(47)$ & $0.0401(45)$\\ \hline
    
    Classical light&&&&&\\
    
   $\bar{n}$  & $A_1 A_2$ & $A_2 A_3$ & $A_3 A_4$ & $A_4 A_5$ & $A_5 A_1$ \\
    \hline
0.10 & 1.0102(131) & 1.0115(126) & 1.0161(144) & 1.0115(136) & 1.0132(129) \\
0.40 & 1.0096(35) & 1.0086(32) & 1.0104(52) & 1.0090(31) & 1.0092(31) \\
0.72 & 1.0093(17) & 1.0147(23) & 1.0095(18) & 1.0094(17) & 1.0088(17) \\
0.99 & 1.0139(15) & 1.0105(11) & 1.0117(13) & 1.0097(12) & 1.0105(10) \\
1.24 & 1.0093(8) & 1.0091(10) & 1.0099(9) & 1.0099(9) & 1.0092(7) \\
1.57 & 1.0095(7) & 1.0094(6) & 1.0103(6) & 1.0095(7) & 1.0093(6) \\
1.84 & 1.0093(5) & 1.0110(6) & 1.0106(5) & 1.0105(5) & 1.0103(7)
   \end{tabular}
\end{ruledtabular}
    \end{table}
    
\par
\textit{The correlation of the outcomes}.---As clarified in the main text, the correlations between the measurement outcomes play a central role in the violation of the KCBS inequality. The measured values of correlation function $g_{i,i+1}$ [defined in Eq. (2) in the main text] between the outcomes $A_i$ and $A_{i+1}$ in our experiment are summarized in Table \ref{tab:g} for single photons and classical light fields. As we expected, the measured correlations in the single photon experiment are nearly zero. The joint probability term $P(A_i=-1,A_{i+1}=-1)$ thus can be ignored compared with the marginal probabilities $P(A_i=-1)$. In contrast, the measured correlations $g_{i,i+1}$ in the classical light experiments are about 1, which testify the outcomes are uncorrelated. The measured values are slightly larger than 1 due to the intensity fluctuations in the light waves \cite{book:14435}. 
\par
When the light intensity is low, the correlation function $g_{i,i+1}$ defined in the main text can be approximately recast into
\begin{equation}
g_{i,i+1}=\frac{P(A_i=-1,A_{i+1}=-1)}{P(A_i=-1)P(A_{i+1}=-1)}\approx\frac{\langle I_i I_{i+1}\rangle}{\langle I_i\rangle\langle I_{i+1}\rangle},
\end{equation}
where $I_i$ represents the intensity of the light in the port $i$ (which corresponds to the observable $A_i$ in the experiment). This formulas is very similar to the second order coherence $g^{(2)}(0)$ of the light field, which is a witness of non-classicality of light fields and used to quantify the photon antibunching. In practice, the violation of non-contextual inequalities does not solely depend on $g^{(2)}$ of the state, but also depends on the specific photon number distribution (determined by the average photon number for a specific kind of number distribution, e.g., coherent states, thermal states). As we shown in the main text, the violation of the NC inequality with a linear optical setup is related to the photon statistics of the input light field. In the following, we establish the connection between the input state and the violations of non-contextual inequality in the Glauber--Sudarshan $P$ representation.
\par
\textit{Non-classicality and contextuality in Glauber--Sudarshan $P$ representation}.---We start with the mapping of the linear optical network for a coherent state $|\alpha \rangle$ as input in one mode and vacuum state $|0\rangle$ in the other two modes,
\begin{equation}
|0\rangle_{0}|0\rangle_{1}|\alpha\rangle_{2}\mapsto \left|\frac{1}{\sqrt[4]{5}}\alpha\right\rangle_{A_i}\left|\frac{1}{\sqrt[4]{5}}\alpha\right\rangle_{A_{i+1}}\left|\left(1-\frac{2}{\sqrt{5}}\right)^{1/2}\alpha\right\rangle_{A_\text{anc}}.
\end{equation}
Then let us consider a general input state in mode 2 instead of a coherent state. The input field can be represented as $\hat{\rho}_{\text{in}}=|0\rangle_0\langle 0|\otimes|0\rangle_1\langle 0|\otimes\hat{\rho}$, where the the Glauber--Sudarshan $P$ function of $\hat{\rho}$ is denoted by $P(\alpha)$. The Glauber--Sudarshan $P$ representation of the input field can be written as:
\begin{equation}
\hat{\rho}_{\text{in}}=\int d^2\gamma \int d^2\beta \int d^2\alpha \ \delta (\gamma) \delta(\beta) P(\alpha) \ |\gamma\rangle_0\langle\gamma|\otimes|\beta\rangle_1\langle\beta|\otimes|\alpha\rangle_2\langle\alpha|.
\end{equation}
And the output field of the optical network can be represented as
\begin{equation}
\hat{\rho}_{\text{out}}=\int  d^2\alpha \ P(\alpha) \ |\alpha/\sqrt[4]{5}\rangle_{A_i} \langle\alpha/\sqrt[4]{5}|\otimes|\alpha/\sqrt[4]{5}\rangle_{A_{i+1}}\langle\alpha/\sqrt[4]{5}|\otimes|\sqrt{1-2/\sqrt{5}}\alpha\rangle_{A_{\text{anc}}}\langle\sqrt{1-2/\sqrt{5}}\alpha|.
\end{equation}
If the value of $\beta$ for a coherent state $|\alpha\rangle$ as input is denoted as $\beta(\alpha |E_j)=\sum_i M_i\left(\alpha |E_j\right)$ under $E_j$, generally the value of $\beta$ for the state $\hat{\rho}_{\text{in}}$ is
\begin{equation}
\beta_{E_j}=\int  d^2\alpha \ P(\alpha) \ \beta(\alpha |E_j).
\label{eq:pe3}
\end{equation}
In our case, applying the operator $\hat{O}=|0\rangle_{A_i}\langle 0|\otimes|0\rangle_{A_{i+1}}\langle 0|+(\openone-|0\rangle_{A_{i}}\langle 0|)\otimes(\openone-|0\rangle_{A_{i+1}}\langle 0|)-(\openone-|0\rangle_{A_{i}}\langle 0|)\otimes|0\rangle_{A_{i+1}}\langle 0|-|0\rangle_{A_{i}}\langle 0|\otimes(\openone-|0\rangle_{A_{i+1}}\langle 0|)$ on the state, and recalling that the experiment has 5 symmetric contexts, we arrive at the value of $\beta$ under $E_3$
\begin{equation}
\beta_{E_3}=5\times\int d^2\alpha \ P(\alpha) \ \left[1-2\exp\left(-|\alpha|^2/\sqrt{5}\right)\right]^2,
\end{equation}
\begin{equation}
\beta_{E_3}=5\times\int d|\alpha| \ \tilde{P}(|\alpha|) \ \left[1-2\exp\left(-|\alpha|^2/\sqrt{5}\right)\right]^2,
\end{equation}
where $\tilde{P}(|\alpha|)$ represents the qusi-probability distribution $\tilde{P}(|\alpha|)=\int_0^{2\pi}d\phi\ |\alpha|P(\alpha)$, $\phi=\textrm{arg}(\alpha)$~\cite{book:14435}. The photon number distribution of the input state can also be derived by this qusi-probability distribution:
\begin{eqnarray}
\mathcal{P}(n)=&&\text{Tr}(\hat{\rho}_{\text{in}} |n\rangle\langle n|) \nonumber \\
=&&\int d^2 \alpha \ P(\alpha)|\langle n|\alpha\rangle|^2 \nonumber \\
=&&\int d|\alpha|\ \tilde{P}(|\alpha|)\exp(-|\alpha|^2)\frac{|\alpha|^{2n}}{n!}.
\end{eqnarray}
The photon number distribution $\mathcal{P}(n)$ can be represented as an average over Poissonian distributions. To violate the KCBS inequality the $\beta$ value should be less than $-3$, i.e.
\begin{equation}
5\times\int  d^2\alpha \ P(\alpha) \ \left[1-2\exp\left(-|\alpha|^2/\sqrt{5}\right)\right]^2<-3.
\end{equation}
\par
In our case, since the $\beta$ value for a coherent state input always can not violate the inequality in spite of the average photon number [$\text{min}(\beta)=0$ for coherent states], there should be some non-classicality in the P-representation of the input state $P(\alpha)$ if a violation is observed with this input state.
\par
Moreover, if the measurement device has a detection efficiency $\eta$ due to some losses in the setup, the output state should be
\begin{equation}
|0\rangle_{0}|0\rangle_{1}|\alpha\rangle_{2}\mapsto \left|\frac{\sqrt{\eta}}{\sqrt[4]{5}}\alpha\right\rangle_{A_i}\left|\frac{\sqrt{\eta}}{\sqrt[4]{5}}\alpha\right\rangle_{A_{i+1}}\left|\sqrt{\eta}\left(1-\frac{2}{\sqrt{5}}\right)^{1/2}\alpha\right\rangle_{A_\text{anc}},
\label{eq:state}
\end{equation} 
which yields the result
\begin{equation}
\beta_{E_3}=5\times\int  d^2\alpha \ P(\alpha) \ \left[1-2\exp\left(-\eta |\alpha|^2/\sqrt{5}\right)\right]^2.
\end{equation}
\par
In summary, the violation of the NC inequality without post-selection will depend on the non-classicality of the input light field. As an example, we perform numerical simulations for the mixture of single photon state $|1\rangle$, coherent state $|\alpha\rangle$ and thermal state $\rho_{\text{th}}=\sum_n \bar{n}_{\text{th}}^n /(1+\bar{n}_{\text{th}})^{n+1}|n\rangle\langle n|$ to demonstrate different violations of the KCBS inequality. Specifically, we show the simulation results for 2 kinds of optical states:
\par
(i) The mixture of the single photon state $|1\rangle$ and the coherent state $|\alpha\rangle$,
\begin{equation}
\rho_1=(1-\lambda)|1\rangle\langle 1|+\lambda|\alpha\rangle\langle\alpha|.
\end{equation}
\par
(ii) The mixture of the single photon state $|1\rangle$ and the thermal state $\rho_{\text{thermal}}$ with average photon number $n_{th}$,
\begin{equation}
\rho_2=(1-\lambda)|1\rangle\langle 1|+\lambda\rho_{\text{th}}.
\end{equation}
\par
The simulation results are shown in Fig. \ref{fig:mix}. As we expected, violating the KCBS inequality is a sufficient but not necessary condition for the negativity of $P(\alpha)$. The $P$ functions of the two states generally possess negativity since they have a single photon state in the mixture, but violating the inequality demands a higher ratio of single photon state in the mixture.
\begin{figure}
\centering
\addtocounter{figure}{1}
\subfigure{\includegraphics[width=0.45\linewidth]{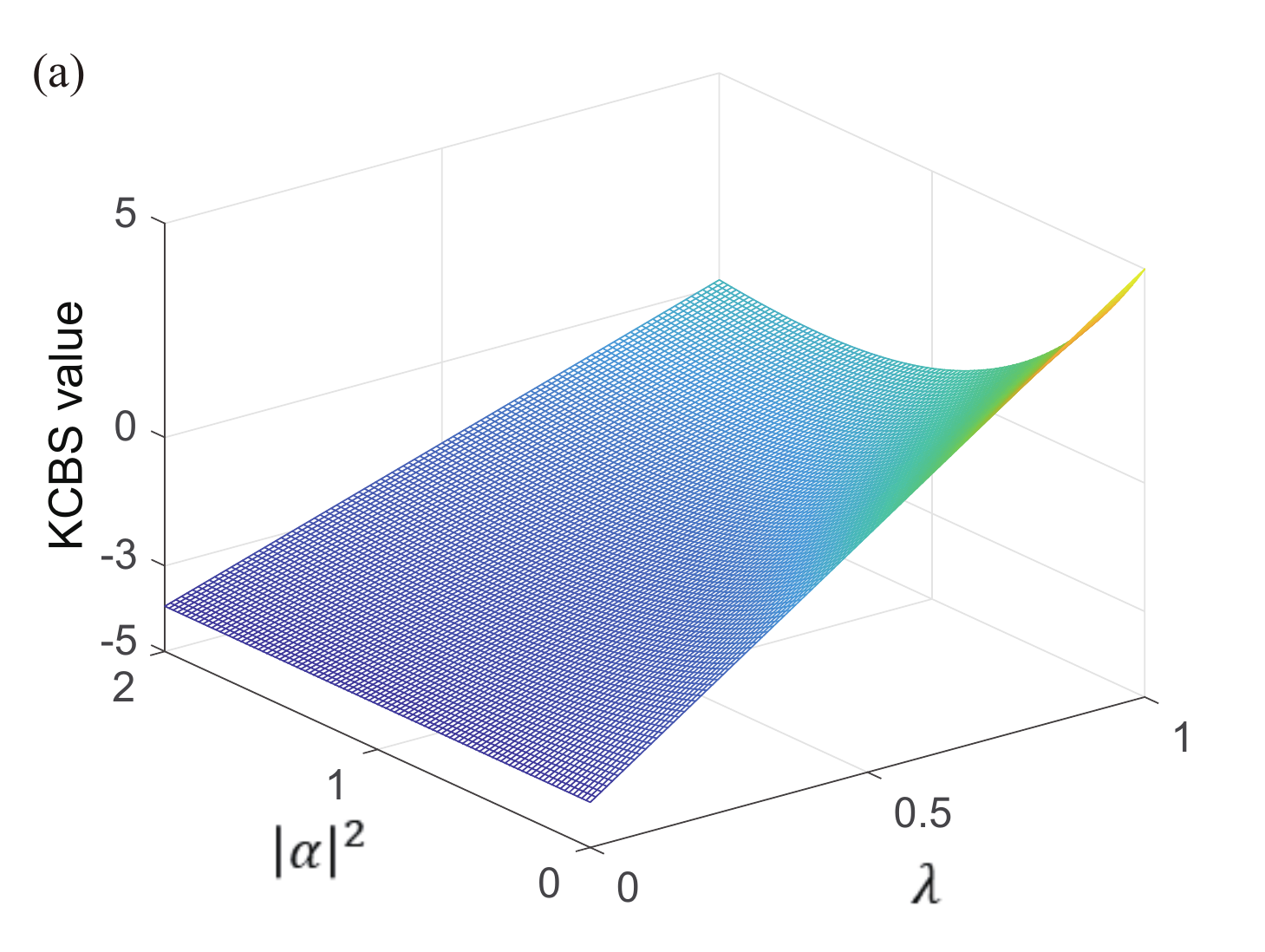}\label{fig:mix:a}}
\quad
\subfigure{\includegraphics[width=0.45\linewidth]{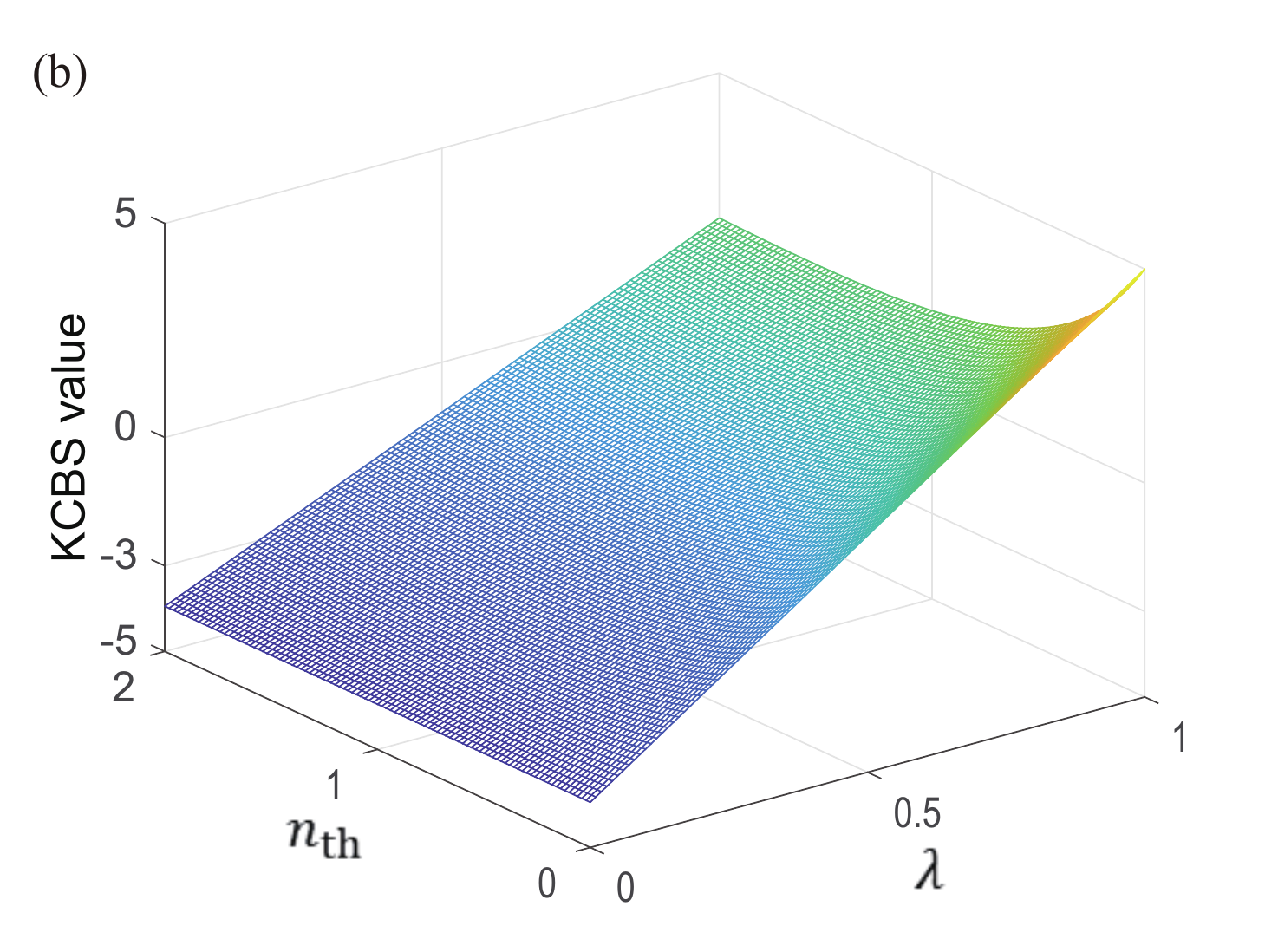}\label{fig:mix:b}}
\addtocounter{figure}{-1}
\caption{\label{fig:mix} The simulation results of different optical states: (a) the mixture of single photon state $|1\rangle$ and coherent state $|\alpha\rangle$; (b) the mixture of single photon state $|1\rangle$ and thermal state $\rho_{\text{th}}$.}
\end{figure}
\par
On the other hand, if we apply post-selection on the detection part, the value of $\beta$ for the measurement event $E_1$ with perfect detection can be written as:
\begin{equation}
\beta_{E_1}=5\times\int  d^2\alpha \ P(\alpha) \ \frac{\exp\left(-2|\alpha|^2/\sqrt{5}\right)-2\exp\left[-\left(\sqrt{5}-1\right)|\alpha|^2/\sqrt{5}\right]+\exp(-|\alpha|^2)} {\exp\left(-2|\alpha|^2/\sqrt{5}\right)+2\exp\left[-\left(\sqrt{5}-1\right)|\alpha|^2/\sqrt{5}\right]-3\exp(-|\alpha|^2)}.
\label{eq:pe1}
\end{equation}
Since $\beta (\alpha |E_1)\leq 5-4\sqrt{5}$, the inequality can be violated under this condition even with a classical $P(\alpha)$ as input. For $E_2$ we have

\begin{equation}
\beta_{E_2}=5\times\int  d^2\alpha \ P(\alpha) \ \frac{1-\exp\left(-|\alpha|^2\right)+4\exp\left(-2|\alpha|^2/\sqrt{5}\right)-4\exp\left(-|\alpha|^2/\sqrt{5}\right)} {1-\exp\left(-|\alpha|^2\right)}.
\label{eq:pe2}
\end{equation}
The equation can also be represented in terms of post-selection probability $P(E_2)$,
\begin{equation}
\beta_{E_2}=5\times\int  d^2\alpha \ P(\alpha) \ \left[1+\frac{\beta \left(\alpha |E_3\right)}{P(E_2)}-\frac{1}{P(E_2)}\right].
\end{equation}
Substituting the $P$ function of the single photon state $P(\alpha)=\left(1+\frac{1}{4}\Delta \right)\delta(\alpha)$ into Eq. (\ref{eq:pe3}), one can find that the KCBS value of $\beta(\alpha|E_1)$ under the limit $\alpha \to 0$ is equal to the value of single photon state under $E_3$. Similarly, consider the state 
\begin{equation}
|\mu \rangle=\sum_{n=1}^{\infty}\frac{\exp(-|\alpha_0|^2/2)}{[1-\exp(-|\alpha_0|^2)]^{1/2}}\frac{\alpha_0^n}{(n!)^{1/2}}|n\rangle
\end{equation}
which is equivalent to a coherent state with vacuum component removed. The state possesses a $P$ function 
$\displaystyle{
P(\alpha)=\frac{1}{1-\exp(-|\alpha_0|^2)}\delta(\alpha-\alpha_0)-\frac{\exp(-|\alpha_0|^2)}{1-\exp(-|\alpha_0|^2)}\delta(\alpha)}$,
and the KCBS value of $\beta(\alpha_0|E_2)$ is equal to the value for the state $|\mu \rangle$ as input under $E_3$.
\par
For the single photon state $|1\rangle$ with $P(\alpha)=\left(1+\frac{1}{4}\Delta \right)\delta(\alpha)$ as the input, it can be found that the violation is the same under the three definitions $E_1, E_2$ and $E_3$, i.e. $\beta_{E_1}=\beta_{E_2}=\beta_{E_3}=5-4\sqrt{5}$ according to Eq. (\ref{eq:pe3}).

\section{Experimental data}
\begin{table}[h]
\centering
\caption{The registered measurement results for single photon experiment. Here $C_{a,(b,c),T}$ denote the coincidence count rate in 1.4s between the detectors $a,(b,c)$ and the trigger detector (the idler mode in SPDC), and $N_T$ denotes the corresponding count rate in the trigger detector. The values in the brackets represent standard uncertainties given by 150 repetitions of the experiment and data collection.}
\label{raw-single}
\begin{ruledtabular}
\begin{tabular}{rrrrrrrrr}
              & $C_{1,T}$  & $C_{2,T}$  & $C_{3,T}$  & $C_{1,2,T}$ & $C_{1,3,T}$ & $C_{2,3,T}$ & $C_{1,2,3,T}$ & $N_{T}$      \\ \hline
$\{A_1,A_2\}$ & 25347(193) & 25075(178) & 6082(84)   & 72(9)       & 19(4)       & 19(5)       & 0(0)          & 346798(1640) \\ 
$\{A_2,A_3\}$ & 26569(202) & 5696(86)   & 25698(194) & 20(4)       & 77(9)       & 19(4)       & 0(0)          & 356275(1352) \\ 
$\{A_3,A_4\}$ & 6494(80)   & 24918(178) & 26345(211) & 20(5)       & 21(5)       & 75(8)       & 0(0)          & 360549(1469) \\ 
$\{A_4,A_5\}$ & 27041(225) & 25313(198) & 6779(86)   & 77(9)       & 24(5)       & 23(5)       & 0(0)          & 363776(2026) \\ 
$\{A_5,A_1\}$ & 27349(207) & 26333(178) & 6244(91)   & 79(9)       & 22(5)       & 21(5)       & 0(0)          & 366636(1353) \\ 
\end{tabular}
\end{ruledtabular}
\end{table}

\begin{table}[h]
\centering
\caption{The registered measurement results for classical light filed inputs. Here $C_{a,(b,c),T}$ denote the coincidence count rate in 1.4s between the detectors $a,(b,c)$ and the trigger signal (the synchronizing signal of the pulse picker), and $N_T$ denotes the corresponding count rate in the trigger signal. The values in the brackets represent standard uncertainties given by 150 repetitions of the experiment and data collection.}
\label{raw-classical}
\begin{ruledtabular}
\begin{tabular}{rrrrrrrrr}
$\bar{n}=0.10$  & $C_{1,T}$      & $C_{2,T}$      & $C_{3,T}$     & $C_{1,2,T}$   & $C_{1,3,T}$   & $C_{2,3,T}$  & $C_{1,2,3,T}$ & $N_{T}$    \\ \hline
$\{A_1,A_2\}$  & 119146(656)    & 118814(674)    & 29801(212)    & 5519(91)      & 1390(41)      & 1390(40)     & 65(8)         & 2591148(1) \\
$\{A_2,A_3\}$  & 120200(941)    & 26983(281)     & 120803(1048)  & 1268(42)      & 5669(112)     & 1277(45)     & 60(8)         & 2591149(0) \\
$\{A_3,A_4\}$  & 29812(262)     & 114535(863)    & 120642(971)   & 1341(41)      & 1405(38)      & 5416(91)     & 62(8)         & 2591154(0) \\
$\{A_4,A_5\}$  & 119672(1042)   & 112374(1281)   & 30486(456)    & 5250(119)     & 1429(45)      & 1344(48)     & 63(8)         & 2591153(1) \\
$\{A_5,A_1\}$  & 118994(943)    & 114533(951)    & 26860(288)    & 5329(110)     & 1259(44)      & 1217(43)     & 57(7)         & 2591152(0) \\ \hline
               &                &                &               &               &               &              &               &            \\
$\bar{n}=0.40$
               & $C_{1,T}$      & $C_{2,T}$      & $C_{3,T}$     & $C_{1,2,T}$   & $C_{1,3,T}$   & $C_{2,3,T}$  & $C_{1,2,3,T}$ & $N_{T}$    \\ \hline
$\{A_1,A_2\}$  & 421355(3326)   & 418210(3245)   & 109382(1042)  & 68666(1068)   & 17994(351)    & 17886(323)   & 2948(93)      & 2591147(0) \\
$\{A_2,A_3\}$  & 433901(1231)   & 103593(386)    & 432928(1220)  & 17508(150)    & 73118(439)    & 17511(154)   & 2964(58)      & 2591147(0) \\
$\{A_3,A_4\}$  & 115941(1474)   & 405581(2188)   & 427735(2393)  & 18309(298)    & 19348(344)    & 67625(763)   & 3071(82)      & 2591145(0) \\
$\{A_4,A_5\}$  & 431660(4489)   & 402893(3439)   & 114514(1112)  & 67727(1263)   & 19314(362)    & 18032(325)   & 3050(90)      & 2591145(0) \\
$\{A_5,A_1\}$  & 424048(4667)   & 410394(3930)   & 104782(1548)  & 67783(1055)   & 17333(353)    & 16789(386)   & 2779(93)      & 2591146(1) \\ \hline
               &                &                &               &               &               &              &               &            \\
$\bar{n}=0.72$
               & $C_{1,T}$      & $C_{2,T}$      & $C_{3,T}$     & $C_{1,2,T}$   & $C_{1,3,T}$   & $C_{2,3,T}$  & $C_{1,2,3,T}$ & $N_{T}$    \\ \hline
$\{A_1,A_2\}$  & 717113(7035)   & 690496(6908)   & 194970(2247)  & 192895(3832)  & 54581(1185)   & 52589(1121)  & 14782(465)    & 2591145(0) \\
$\{A_2,A_3\}$  & 707364(11448)  & 185247(3448)   & 680115(11717) & 51256(1715)   & 188436(5988)  & 49489(1696)  & 13765(675)    & 2591144(1) \\
$\{A_3,A_4\}$  & 210232(1300)   & 697006(2022)   & 732012(1846)  & 57056(459)    & 59952(515)    & 198775(1090) & 16362(194)    & 2591144(0) \\
$\{A_4,A_5\}$  & 719289(5133)   & 685378(4151)   & 199189(1474)  & 192063(2528)  & 55986(766)    & 53349(709)   & 15053(311)    & 2591142(0) \\
$\{A_5,A_1\}$  & 731302(6914)   & 712236(4699)   & 196599(1824)  & 202803(3202)  & 56016(988)    & 54567(824)   & 15611(375)    & 2591142(0) \\ \hline
               &                &                &               &               &               &              &               &            \\
$\bar{n}=0.99$ 
               & $C_{1,T}$      & $C_{2,T}$      & $C_{3,T}$     & $C_{1,2,T}$   & $C_{1,3,T}$   & $C_{2,3,T}$  & $C_{1,2,3,T}$ & $N_{T}$    \\ \hline
$\{A_1,A_2\}$  & 897217(8187)   & 861074(7947)   & 251446(2828)  & 302309(5271)  & 88699(1711)   & 85192(1685)  & 30156(848)    & 2591147(1) \\
$\{A_2,A_3\}$  & 959875(2193)   & 265437(829)    & 924061(2337)  & 99344(565)    & 345923(1643)  & 95879(561)   & 36051(329)    & 2591149(0) \\
$\{A_3,A_4\}$  & 278189(1557)   & 926253(3816)   & 925607(4612)  & 100397(981)   & 100443(1007)  & 334755(2851) & 36496(522)    & 2591149(0) \\
$\{A_4,A_5\}$  & 946736(5764)   & 923119(5554)   & 273625(2350)  & 340558(3975)  & 101080(1364)  & 98725(1274)  & 36627(681)    & 2591148(1) \\
$\{A_5,A_1\}$  & 951452(5331)   & 926865(6464)   & 279351(1825)  & 343924(4237)  & 103876(814)   & 101411(823)  & 37861(491)    & 2591147(0) \\ \hline
               &                &                &               &               &               &              &               &            \\
$\bar{n}=1.24$ 
               & $C_{1,T}$      & $C_{2,T}$      & $C_{3,T}$     & $C_{1,2,T}$   & $C_{1,3,T}$   & $C_{2,3,T}$  & $C_{1,2,3,T}$ & $N_{T}$    \\ \hline
$\{A_1,A_2\}$  & 1115288(5367)  & 1081923(5058)  & 326521(2343)  & 470045(4425)  & 141901(1621)  & 137749(1525) & 60177(958)    & 2591146(0) \\
$\{A_2,A_3\}$  & 1126739(5336)  & 314891(2065)   & 1101507(6182) & 138200(1550)  & 483343(4754)  & 135289(1568) & 59669(963)    & 2591145(0) \\
$\{A_3,A_4\}$  & 338075(1071)   & 1078697(1676)  & 1106430(1658) & 141996(594)   & 145671(629)   & 465190(1360) & 61561(365)    & 2591145(0) \\
$\{A_4,A_5\}$  & 1106730(6316)  & 1067832(7400)  & 335458(2607)  & 460636(5708)  & 144953(1823)  & 140054(2003) & 60772(1177)   & 2591144(0) \\
$\{A_5,A_1\}$  & 1115093(2525)  & 1090014(4168)  & 335552(4281)  & 473394(2716)  & 145785(1779)  & 142681(1584) & 62308(705)    & 2591144(0) \\ \hline
               &                &                &               &               &               &              &               &            \\
$\bar{n}=1.57$ 
               & $C_{1,T}$      & $C_{2,T}$      & $C_{3,T}$     & $C_{1,2,T}$   & $C_{1,3,T}$   & $C_{2,3,T}$  & $C_{1,2,3,T}$ & $N_{T}$    \\ \hline
$\{A_1,A_2\}$  & 1318906(5593)  & 1291769(5611)  & 428766(2478)  & 663785(5796)  & 220573(2247)  & 216089(2230) & 111736(1647)  & 2591216(0) \\
$\{A_2,A_3\}$  & 1346500(3073)  & 397968(1320)   & 1345865(3122) & 208695(1161)  & 705972(3099)  & 208919(1165) & 110158(890)   & 2591215(1) \\
$\{A_3,A_4\}$  & 415002(2674)   & 1240955(5240)  & 1303059(6136) & 200562(2146)  & 210617(2231)  & 630461(5533) & 102429(1536)  & 2591208(0) \\
$\{A_4,A_5\}$  & 1280196(4707)  & 1230235(4300)  & 426232(2684)  & 613613(4297)  & 212748(1979)  & 204461(1876) & 102607(1258)  & 2591209(0) \\
$\{A_5,A_1\}$  & 1324944(5144)  & 1311555(4218)  & 379163(2571)  & 676899(4759)  & 195756(2029)  & 193872(1848) & 100637(1352)  & 2591210(0) \\  \hline
               &                &                &               &               &               &              &               &            \\
$\bar{n}=1.84$
               & $C_{1,T}$      & $C_{2,T}$      & $C_{3,T}$     & $C_{1,2,T}$   & $C_{1,3,T}$   & $C_{2,3,T}$  & $C_{1,2,3,T}$ & $N_{T}$    \\ \hline
$\{A_1,A_2\}$  & 1455626(2947)  & 1424706(2985)  & 492345(1515)  & 807816(3273)  & 279096(1416)  & 273356(1438) & 155817(1146)  & 2591222(1) \\
$\{A_2,A_3\}$  & 1449516(9606)  & 452415(4344)   & 1439459(9698) & 255813(4169)  & 814156(10871) & 254502(4142) & 144729(3354)  & 2591223(0) \\
$\{A_3,A_4\}$  & 480699(1714)   & 1425190(4068)  & 1465853(3883) & 266827(1591)  & 274549(1591)  & 814810(4481) & 153422(1311)  & 2591224(0) \\
$\{A_4,A_5\}$  & 1463814(11009) & 1424103(10769) & 511443(2737)  & 813015(12255) & 292407(3499)  & 284713(3440) & 163604(3188)  & 2591225(0) \\
$\{A_5,A_1\}$  & 1476570(7297)  & 1468857(8392)  & 454731(3931)  & 845609(8594)  & 262060(3331)  & 260935(3443) & 151153(2628)  & 2591226(0)
\end{tabular}
\end{ruledtabular}
\end{table}
\end{widetext}

\begin{thebibliography}{47}%
\makeatletter
\providecommand \@ifxundefined [1]{%
 \@ifx{#1\undefined}
}%
\providecommand \@ifnum [1]{%
 \ifnum #1\expandafter \@firstoftwo
 \else \expandafter \@secondoftwo
 \fi
}%
\providecommand \@ifx [1]{%
 \ifx #1\expandafter \@firstoftwo
 \else \expandafter \@secondoftwo
 \fi
}%
\providecommand \natexlab [1]{#1}%
\providecommand \enquote  [1]{``#1''}%
\providecommand \bibnamefont  [1]{#1}%
\providecommand \bibfnamefont [1]{#1}%
\providecommand \citenamefont [1]{#1}%
\providecommand \href@noop [0]{\@secondoftwo}%
\providecommand \href [0]{\begingroup \@sanitize@url \@href}%
\providecommand \@href[1]{\@@startlink{#1}\@@href}%
\providecommand \@@href[1]{\endgroup#1\@@endlink}%
\providecommand \@sanitize@url [0]{\catcode `\\12\catcode `\$12\catcode
  `\&12\catcode `\#12\catcode `\^12\catcode `\_12\catcode `\%12\relax}%
\providecommand \@@startlink[1]{}%
\providecommand \@@endlink[0]{}%
\providecommand \url  [0]{\begingroup\@sanitize@url \@url }%
\providecommand \@url [1]{\endgroup\@href {#1}{\urlprefix }}%
\providecommand \urlprefix  [0]{URL }%
\providecommand \Eprint [0]{\href }%
\providecommand \doibase [0]{http://dx.doi.org/}%
\providecommand \selectlanguage [0]{\@gobble}%
\providecommand \bibinfo  [0]{\@secondoftwo}%
\providecommand \bibfield  [0]{\@secondoftwo}%
\providecommand \translation [1]{[#1]}%
\providecommand \BibitemOpen [0]{}%
\providecommand \bibitemStop [0]{}%
\providecommand \bibitemNoStop [0]{.\EOS\space}%
\providecommand \EOS [0]{\spacefactor3000\relax}%
\providecommand \BibitemShut  [1]{\csname bibitem#1\endcsname}%
\let\auto@bib@innerbib\@empty
\bibitem [{\citenamefont {Kochen}\ and\ \citenamefont
  {Specker}(1967)}]{Kochen1967}%
  \BibitemOpen
  \bibfield  {author} {\bibinfo {author} {\bibfnamefont {S.}~\bibnamefont
  {Kochen}}\ and\ \bibinfo {author} {\bibfnamefont {E.~P.}\ \bibnamefont
  {Specker}},\ }\href {http://www.jstor.org/stable/24902153} {\bibfield
  {journal} {\bibinfo  {journal} {Journal of Mathematics and Mechanics}\
  }\textbf {\bibinfo {volume} {17}},\ \bibinfo {pages} {59} (\bibinfo {year}
  {1967})}\BibitemShut {NoStop}%
\bibitem [{\citenamefont {BELL}(1966)}]{JOHNS.1966}%
  \BibitemOpen
  \bibfield  {author} {\bibinfo {author} {\bibfnamefont {J.~S.}\ \bibnamefont
  {BELL}},\ }\href {\doibase 10.1103/RevModPhys.38.447} {\bibfield  {journal}
  {\bibinfo  {journal} {Rev. Mod. Phys.}\ }\textbf {\bibinfo {volume} {38}},\
  \bibinfo {pages} {447} (\bibinfo {year} {1966})}\BibitemShut {NoStop}%
\bibitem [{\citenamefont {Lapkiewicz}\ \emph {et~al.}(2011)\citenamefont
  {Lapkiewicz}, \citenamefont {Li}, \citenamefont {Schaeff}, \citenamefont
  {Langford}, \citenamefont {Ramelow}, \citenamefont {Wie{\'{s}}niak},\ and\
  \citenamefont {Zeilinger}}]{Lapkiewicz_2011}%
  \BibitemOpen
  \bibfield  {author} {\bibinfo {author} {\bibfnamefont {R.}~\bibnamefont
  {Lapkiewicz}}, \bibinfo {author} {\bibfnamefont {P.}~\bibnamefont {Li}},
  \bibinfo {author} {\bibfnamefont {C.}~\bibnamefont {Schaeff}}, \bibinfo
  {author} {\bibfnamefont {N.~K.}\ \bibnamefont {Langford}}, \bibinfo {author}
  {\bibfnamefont {S.}~\bibnamefont {Ramelow}}, \bibinfo {author} {\bibfnamefont
  {M.}~\bibnamefont {Wie{\'{s}}niak}}, \ and\ \bibinfo {author} {\bibfnamefont
  {A.}~\bibnamefont {Zeilinger}},\ }\href {\doibase 10.1038/nature10119}
  {\bibfield  {journal} {\bibinfo  {journal} {Nature}\ }\textbf {\bibinfo
  {volume} {474}},\ \bibinfo {pages} {490} (\bibinfo {year}
  {2011})}\BibitemShut {NoStop}%
\bibitem [{\citenamefont {Kirchmair}\ \emph {et~al.}(2009)\citenamefont
  {Kirchmair}, \citenamefont {Zahringer}, \citenamefont {Gerritsma},
  \citenamefont {Kleinmann}, \citenamefont {Guhne}, \citenamefont {Cabello},
  \citenamefont {Blatt},\ and\ \citenamefont {Roos}}]{Kirchmair2009}%
  \BibitemOpen
  \bibfield  {author} {\bibinfo {author} {\bibfnamefont {G.}~\bibnamefont
  {Kirchmair}}, \bibinfo {author} {\bibfnamefont {F.}~\bibnamefont
  {Zahringer}}, \bibinfo {author} {\bibfnamefont {R.}~\bibnamefont
  {Gerritsma}}, \bibinfo {author} {\bibfnamefont {M.}~\bibnamefont
  {Kleinmann}}, \bibinfo {author} {\bibfnamefont {O.}~\bibnamefont {Guhne}},
  \bibinfo {author} {\bibfnamefont {A.}~\bibnamefont {Cabello}}, \bibinfo
  {author} {\bibfnamefont {R.}~\bibnamefont {Blatt}}, \ and\ \bibinfo {author}
  {\bibfnamefont {C.~F.}\ \bibnamefont {Roos}},\ }\href
  {http://dx.doi.org/10.1038/nature08172} {\bibfield  {journal} {\bibinfo
  {journal} {Nature}\ }\textbf {\bibinfo {volume} {460}},\ \bibinfo {pages}
  {494} (\bibinfo {year} {2009})}\BibitemShut {NoStop}%
\bibitem [{\citenamefont {Zu}\ \emph {et~al.}(2012)\citenamefont {Zu},
  \citenamefont {Wang}, \citenamefont {Deng}, \citenamefont {Chang},
  \citenamefont {Liu}, \citenamefont {Hou}, \citenamefont {Yang},\ and\
  \citenamefont {Duan}}]{PhysRevLett.109.150401}%
  \BibitemOpen
  \bibfield  {author} {\bibinfo {author} {\bibfnamefont {C.}~\bibnamefont
  {Zu}}, \bibinfo {author} {\bibfnamefont {Y.-X.}\ \bibnamefont {Wang}},
  \bibinfo {author} {\bibfnamefont {D.-L.}\ \bibnamefont {Deng}}, \bibinfo
  {author} {\bibfnamefont {X.-Y.}\ \bibnamefont {Chang}}, \bibinfo {author}
  {\bibfnamefont {K.}~\bibnamefont {Liu}}, \bibinfo {author} {\bibfnamefont
  {P.-Y.}\ \bibnamefont {Hou}}, \bibinfo {author} {\bibfnamefont {H.-X.}\
  \bibnamefont {Yang}}, \ and\ \bibinfo {author} {\bibfnamefont {L.-M.}\
  \bibnamefont {Duan}},\ }\href {\doibase 10.1103/PhysRevLett.109.150401}
  {\bibfield  {journal} {\bibinfo  {journal} {Phys. Rev. Lett.}\ }\textbf
  {\bibinfo {volume} {109}},\ \bibinfo {pages} {150401} (\bibinfo {year}
  {2012})}\BibitemShut {NoStop}%
\bibitem [{\citenamefont {Huang}\ \emph {et~al.}(2003)\citenamefont {Huang},
  \citenamefont {Li}, \citenamefont {Zhang}, \citenamefont {Pan},\ and\
  \citenamefont {Guo}}]{PhysRevLett.90.250401}%
  \BibitemOpen
  \bibfield  {author} {\bibinfo {author} {\bibfnamefont {Y.-F.}\ \bibnamefont
  {Huang}}, \bibinfo {author} {\bibfnamefont {C.-F.}\ \bibnamefont {Li}},
  \bibinfo {author} {\bibfnamefont {Y.-S.}\ \bibnamefont {Zhang}}, \bibinfo
  {author} {\bibfnamefont {J.-W.}\ \bibnamefont {Pan}}, \ and\ \bibinfo
  {author} {\bibfnamefont {G.-C.}\ \bibnamefont {Guo}},\ }\href {\doibase
  10.1103/PhysRevLett.90.250401} {\bibfield  {journal} {\bibinfo  {journal}
  {Phys. Rev. Lett.}\ }\textbf {\bibinfo {volume} {90}},\ \bibinfo {pages}
  {250401} (\bibinfo {year} {2003})}\BibitemShut {NoStop}%
\bibitem [{\citenamefont {Hu}\ \emph {et~al.}(2016)\citenamefont {Hu},
  \citenamefont {Chen}, \citenamefont {Liu}, \citenamefont {Guo}, \citenamefont
  {Huang}, \citenamefont {Zhou}, \citenamefont {Han}, \citenamefont {Li},\ and\
  \citenamefont {Guo}}]{PhysRevLett.117.170403}%
  \BibitemOpen
  \bibfield  {author} {\bibinfo {author} {\bibfnamefont {X.-M.}\ \bibnamefont
  {Hu}}, \bibinfo {author} {\bibfnamefont {J.-S.}\ \bibnamefont {Chen}},
  \bibinfo {author} {\bibfnamefont {B.-H.}\ \bibnamefont {Liu}}, \bibinfo
  {author} {\bibfnamefont {Y.}~\bibnamefont {Guo}}, \bibinfo {author}
  {\bibfnamefont {Y.-F.}\ \bibnamefont {Huang}}, \bibinfo {author}
  {\bibfnamefont {Z.-Q.}\ \bibnamefont {Zhou}}, \bibinfo {author}
  {\bibfnamefont {Y.-J.}\ \bibnamefont {Han}}, \bibinfo {author} {\bibfnamefont
  {C.-F.}\ \bibnamefont {Li}}, \ and\ \bibinfo {author} {\bibfnamefont {G.-C.}\
  \bibnamefont {Guo}},\ }\href {\doibase 10.1103/PhysRevLett.117.170403}
  {\bibfield  {journal} {\bibinfo  {journal} {Phys. Rev. Lett.}\ }\textbf
  {\bibinfo {volume} {117}},\ \bibinfo {pages} {170403} (\bibinfo {year}
  {2016})}\BibitemShut {NoStop}%
\bibitem [{\citenamefont {Howard}\ \emph {et~al.}(2014)\citenamefont {Howard},
  \citenamefont {Wallman}, \citenamefont {Veitch},\ and\ \citenamefont
  {Emerson}}]{Howard2014}%
  \BibitemOpen
  \bibfield  {author} {\bibinfo {author} {\bibfnamefont {M.}~\bibnamefont
  {Howard}}, \bibinfo {author} {\bibfnamefont {J.}~\bibnamefont {Wallman}},
  \bibinfo {author} {\bibfnamefont {V.}~\bibnamefont {Veitch}}, \ and\ \bibinfo
  {author} {\bibfnamefont {J.}~\bibnamefont {Emerson}},\ }\href
  {http://dx.doi.org/10.1038/nature13460} {\bibfield  {journal} {\bibinfo
  {journal} {Nature}\ }\textbf {\bibinfo {volume} {510}},\ \bibinfo {pages}
  {351} (\bibinfo {year} {2014})}\BibitemShut {NoStop}%
\bibitem [{\citenamefont {Delfosse}\ \emph {et~al.}(2015)\citenamefont
  {Delfosse}, \citenamefont {Allard~Guerin}, \citenamefont {Bian},\ and\
  \citenamefont {Raussendorf}}]{PhysRevX.5.021003}%
  \BibitemOpen
  \bibfield  {author} {\bibinfo {author} {\bibfnamefont {N.}~\bibnamefont
  {Delfosse}}, \bibinfo {author} {\bibfnamefont {P.}~\bibnamefont
  {Allard~Guerin}}, \bibinfo {author} {\bibfnamefont {J.}~\bibnamefont {Bian}},
  \ and\ \bibinfo {author} {\bibfnamefont {R.}~\bibnamefont {Raussendorf}},\
  }\href {\doibase 10.1103/PhysRevX.5.021003} {\bibfield  {journal} {\bibinfo
  {journal} {Phys. Rev. X}\ }\textbf {\bibinfo {volume} {5}},\ \bibinfo {pages}
  {021003} (\bibinfo {year} {2015})}\BibitemShut {NoStop}%
\bibitem [{\citenamefont {Dawkins}\ and\ \citenamefont
  {Howard}(2015)}]{PhysRevLett.115.030501}%
  \BibitemOpen
  \bibfield  {author} {\bibinfo {author} {\bibfnamefont {H.}~\bibnamefont
  {Dawkins}}\ and\ \bibinfo {author} {\bibfnamefont {M.}~\bibnamefont
  {Howard}},\ }\href {\doibase 10.1103/PhysRevLett.115.030501} {\bibfield
  {journal} {\bibinfo  {journal} {Phys. Rev. Lett.}\ }\textbf {\bibinfo
  {volume} {115}},\ \bibinfo {pages} {030501} (\bibinfo {year}
  {2015})}\BibitemShut {NoStop}%
\bibitem [{\citenamefont {Raussendorf}(2013)}]{PhysRevA.88.022322}%
  \BibitemOpen
  \bibfield  {author} {\bibinfo {author} {\bibfnamefont {R.}~\bibnamefont
  {Raussendorf}},\ }\href {\doibase 10.1103/PhysRevA.88.022322} {\bibfield
  {journal} {\bibinfo  {journal} {Phys. Rev. A}\ }\textbf {\bibinfo {volume}
  {88}},\ \bibinfo {pages} {022322} (\bibinfo {year} {2013})}\BibitemShut
  {NoStop}%
\bibitem [{\citenamefont {Bermejo-Vega}\ \emph {et~al.}(2017)\citenamefont
  {Bermejo-Vega}, \citenamefont {Delfosse}, \citenamefont {Browne},
  \citenamefont {Okay},\ and\ \citenamefont
  {Raussendorf}}]{PhysRevLett.119.120505}%
  \BibitemOpen
  \bibfield  {author} {\bibinfo {author} {\bibfnamefont {J.}~\bibnamefont
  {Bermejo-Vega}}, \bibinfo {author} {\bibfnamefont {N.}~\bibnamefont
  {Delfosse}}, \bibinfo {author} {\bibfnamefont {D.~E.}\ \bibnamefont
  {Browne}}, \bibinfo {author} {\bibfnamefont {C.}~\bibnamefont {Okay}}, \ and\
  \bibinfo {author} {\bibfnamefont {R.}~\bibnamefont {Raussendorf}},\ }\href
  {\doibase 10.1103/PhysRevLett.119.120505} {\bibfield  {journal} {\bibinfo
  {journal} {Phys. Rev. Lett.}\ }\textbf {\bibinfo {volume} {119}},\ \bibinfo
  {pages} {120505} (\bibinfo {year} {2017})}\BibitemShut {NoStop}%
\bibitem [{\citenamefont {Nagata}(2005)}]{PhysRevA.72.012325}%
  \BibitemOpen
  \bibfield  {author} {\bibinfo {author} {\bibfnamefont {K.}~\bibnamefont
  {Nagata}},\ }\href {\doibase 10.1103/PhysRevA.72.012325} {\bibfield
  {journal} {\bibinfo  {journal} {Phys. Rev. A}\ }\textbf {\bibinfo {volume}
  {72}},\ \bibinfo {pages} {012325} (\bibinfo {year} {2005})}\BibitemShut
  {NoStop}%
\bibitem [{\citenamefont {Spreeuw}(1998)}]{Spreeuw1998}%
  \BibitemOpen
  \bibfield  {author} {\bibinfo {author} {\bibfnamefont {R.~J.~C.}\
  \bibnamefont {Spreeuw}},\ }\href {\doibase 10.1023/A:1018703709245}
  {\bibfield  {journal} {\bibinfo  {journal} {Foundations of Physics}\ }\textbf
  {\bibinfo {volume} {28}},\ \bibinfo {pages} {361} (\bibinfo {year}
  {1998})}\BibitemShut {NoStop}%
\bibitem [{\citenamefont {T\"oppel}\ \emph {et~al.}(2014)\citenamefont
  {T\"oppel}, \citenamefont {Aiello}, \citenamefont {Marquardt}, \citenamefont
  {Giacobino},\ and\ \citenamefont {Leuchs}}]{T_ppel_2014}%
  \BibitemOpen
  \bibfield  {author} {\bibinfo {author} {\bibfnamefont {F.}~\bibnamefont
  {T\"oppel}}, \bibinfo {author} {\bibfnamefont {A.}~\bibnamefont {Aiello}},
  \bibinfo {author} {\bibfnamefont {C.}~\bibnamefont {Marquardt}}, \bibinfo
  {author} {\bibfnamefont {E.}~\bibnamefont {Giacobino}}, \ and\ \bibinfo
  {author} {\bibfnamefont {G.}~\bibnamefont {Leuchs}},\ }\href {\doibase
  10.1088/1367-2630/16/7/073019} {\bibfield  {journal} {\bibinfo  {journal}
  {New Journal of Physics}\ }\textbf {\bibinfo {volume} {16}},\ \bibinfo
  {pages} {073019} (\bibinfo {year} {2014})}\BibitemShut {NoStop}%
\bibitem [{\citenamefont {Karimi}\ and\ \citenamefont
  {Boyd}(2015)}]{Karimi_2015}%
  \BibitemOpen
  \bibfield  {author} {\bibinfo {author} {\bibfnamefont {E.}~\bibnamefont
  {Karimi}}\ and\ \bibinfo {author} {\bibfnamefont {R.~W.}\ \bibnamefont
  {Boyd}},\ }\href {\doibase 10.1126/science.aad7174} {\bibfield  {journal}
  {\bibinfo  {journal} {Science}\ }\textbf {\bibinfo {volume} {350}},\ \bibinfo
  {pages} {1172} (\bibinfo {year} {2015})}\BibitemShut {NoStop}%
\bibitem [{\citenamefont {Qian}\ \emph {et~al.}(2015)\citenamefont {Qian},
  \citenamefont {Little}, \citenamefont {Howell},\ and\ \citenamefont
  {Eberly}}]{Qian_2015}%
  \BibitemOpen
  \bibfield  {author} {\bibinfo {author} {\bibfnamefont {X.-F.}\ \bibnamefont
  {Qian}}, \bibinfo {author} {\bibfnamefont {B.}~\bibnamefont {Little}},
  \bibinfo {author} {\bibfnamefont {J.~C.}\ \bibnamefont {Howell}}, \ and\
  \bibinfo {author} {\bibfnamefont {J.~H.}\ \bibnamefont {Eberly}},\ }\href
  {\doibase 10.1364/optica.2.000611} {\bibfield  {journal} {\bibinfo  {journal}
  {Optica}\ }\textbf {\bibinfo {volume} {2}},\ \bibinfo {pages} {611} (\bibinfo
  {year} {2015})}\BibitemShut {NoStop}%
\bibitem [{\citenamefont {Kagalwala}\ \emph {et~al.}(2012)\citenamefont
  {Kagalwala}, \citenamefont {Giuseppe}, \citenamefont {Abouraddy},\ and\
  \citenamefont {Saleh}}]{Kagalwala_2012}%
  \BibitemOpen
  \bibfield  {author} {\bibinfo {author} {\bibfnamefont {K.~H.}\ \bibnamefont
  {Kagalwala}}, \bibinfo {author} {\bibfnamefont {G.~D.}\ \bibnamefont
  {Giuseppe}}, \bibinfo {author} {\bibfnamefont {A.~F.}\ \bibnamefont
  {Abouraddy}}, \ and\ \bibinfo {author} {\bibfnamefont {B.~E.~A.}\
  \bibnamefont {Saleh}},\ }\href {\doibase 10.1038/nphoton.2012.312} {\bibfield
   {journal} {\bibinfo  {journal} {Nature Photonics}\ }\textbf {\bibinfo
  {volume} {7}},\ \bibinfo {pages} {72} (\bibinfo {year} {2012})}\BibitemShut
  {NoStop}%
\bibitem [{\citenamefont {Frustaglia}\ \emph {et~al.}(2016)\citenamefont
  {Frustaglia}, \citenamefont {Baltan\'as}, \citenamefont
  {Vel\'azquez-Ahumada}, \citenamefont {Fern\'andez-Prieto}, \citenamefont
  {Lujambio}, \citenamefont {Losada}, \citenamefont {Freire},\ and\
  \citenamefont {Cabello}}]{PhysRevLett.116.250404}%
  \BibitemOpen
  \bibfield  {author} {\bibinfo {author} {\bibfnamefont {D.}~\bibnamefont
  {Frustaglia}}, \bibinfo {author} {\bibfnamefont {J.~P.}\ \bibnamefont
  {Baltan\'as}}, \bibinfo {author} {\bibfnamefont {M.~C.}\ \bibnamefont
  {Vel\'azquez-Ahumada}}, \bibinfo {author} {\bibfnamefont {A.}~\bibnamefont
  {Fern\'andez-Prieto}}, \bibinfo {author} {\bibfnamefont {A.}~\bibnamefont
  {Lujambio}}, \bibinfo {author} {\bibfnamefont {V.}~\bibnamefont {Losada}},
  \bibinfo {author} {\bibfnamefont {M.~J.}\ \bibnamefont {Freire}}, \ and\
  \bibinfo {author} {\bibfnamefont {A.}~\bibnamefont {Cabello}},\ }\href
  {\doibase 10.1103/PhysRevLett.116.250404} {\bibfield  {journal} {\bibinfo
  {journal} {Phys. Rev. Lett.}\ }\textbf {\bibinfo {volume} {116}},\ \bibinfo
  {pages} {250404} (\bibinfo {year} {2016})}\BibitemShut {NoStop}%
\bibitem [{\citenamefont {Li}\ \emph {et~al.}(2017)\citenamefont {Li},
  \citenamefont {Zeng}, \citenamefont {Song},\ and\ \citenamefont
  {Zhang}}]{Li2017}%
  \BibitemOpen
  \bibfield  {author} {\bibinfo {author} {\bibfnamefont {T.}~\bibnamefont
  {Li}}, \bibinfo {author} {\bibfnamefont {Q.}~\bibnamefont {Zeng}}, \bibinfo
  {author} {\bibfnamefont {X.}~\bibnamefont {Song}}, \ and\ \bibinfo {author}
  {\bibfnamefont {X.}~\bibnamefont {Zhang}},\ }\href {\doibase
  10.1038/srep44467} {\bibfield  {journal} {\bibinfo  {journal} {Scientific
  Reports}\ }\textbf {\bibinfo {volume} {7}},\ \bibinfo {pages} {44467}
  (\bibinfo {year} {2017})}\BibitemShut {NoStop}%
\bibitem [{\citenamefont {Markiewicz}\ \emph {et~al.}(2019)\citenamefont
  {Markiewicz}, \citenamefont {Kaszlikowski}, \citenamefont {Kurzy{\'{n}}ski},\
  and\ \citenamefont {W{\'{o}}jcik}}]{Markiewicz2019}%
  \BibitemOpen
  \bibfield  {author} {\bibinfo {author} {\bibfnamefont {M.}~\bibnamefont
  {Markiewicz}}, \bibinfo {author} {\bibfnamefont {D.}~\bibnamefont
  {Kaszlikowski}}, \bibinfo {author} {\bibfnamefont {P.}~\bibnamefont
  {Kurzy{\'{n}}ski}}, \ and\ \bibinfo {author} {\bibfnamefont {A.}~\bibnamefont
  {W{\'{o}}jcik}},\ }\href {https://doi.org/10.1038/s41534-018-0117-8}
  {\bibfield  {journal} {\bibinfo  {journal} {npj Quantum Information}\
  }\textbf {\bibinfo {volume} {5}} (\bibinfo {year} {2019})}\BibitemShut
  {NoStop}%
\bibitem [{\citenamefont {Berg-Johansen}\ \emph {et~al.}(2015)\citenamefont
  {Berg-Johansen}, \citenamefont {T\"{o}ppel}, \citenamefont {Stiller},
  \citenamefont {Banzer}, \citenamefont {Ornigotti}, \citenamefont {Giacobino},
  \citenamefont {Leuchs}, \citenamefont {Aiello},\ and\ \citenamefont
  {Marquardt}}]{Berg-Johansen15}%
  \BibitemOpen
  \bibfield  {author} {\bibinfo {author} {\bibfnamefont {S.}~\bibnamefont
  {Berg-Johansen}}, \bibinfo {author} {\bibfnamefont {F.}~\bibnamefont
  {T\"{o}ppel}}, \bibinfo {author} {\bibfnamefont {B.}~\bibnamefont {Stiller}},
  \bibinfo {author} {\bibfnamefont {P.}~\bibnamefont {Banzer}}, \bibinfo
  {author} {\bibfnamefont {M.}~\bibnamefont {Ornigotti}}, \bibinfo {author}
  {\bibfnamefont {E.}~\bibnamefont {Giacobino}}, \bibinfo {author}
  {\bibfnamefont {G.}~\bibnamefont {Leuchs}}, \bibinfo {author} {\bibfnamefont
  {A.}~\bibnamefont {Aiello}}, \ and\ \bibinfo {author} {\bibfnamefont
  {C.}~\bibnamefont {Marquardt}},\ }\href {\doibase 10.1364/OPTICA.2.000864}
  {\bibfield  {journal} {\bibinfo  {journal} {Optica}\ }\textbf {\bibinfo
  {volume} {2}},\ \bibinfo {pages} {864} (\bibinfo {year} {2015})}\BibitemShut
  {NoStop}%
\bibitem [{\citenamefont {Diego}\ \emph {et~al.}(2016)\citenamefont {Diego},
  \citenamefont {Robert}, \citenamefont {Felix}, \citenamefont {Christian},
  \citenamefont {Markus}, \citenamefont {Matthias}, \citenamefont {Stefan},
  \citenamefont {Michael}, \citenamefont {Andrea}, \citenamefont {Marco},\ and\
  \citenamefont {Alexander}}]{Diego2016}%
  \BibitemOpen
  \bibfield  {author} {\bibinfo {author} {\bibfnamefont {G.-S.}\ \bibnamefont
  {Diego}}, \bibinfo {author} {\bibfnamefont {B.}~\bibnamefont {Robert}},
  \bibinfo {author} {\bibfnamefont {Z.}~\bibnamefont {Felix}}, \bibinfo
  {author} {\bibfnamefont {V.}~\bibnamefont {Christian}}, \bibinfo {author}
  {\bibfnamefont {G.}~\bibnamefont {Markus}}, \bibinfo {author} {\bibfnamefont
  {H.}~\bibnamefont {Matthias}}, \bibinfo {author} {\bibfnamefont
  {N.}~\bibnamefont {Stefan}}, \bibinfo {author} {\bibfnamefont
  {D.}~\bibnamefont {Michael}}, \bibinfo {author} {\bibfnamefont
  {A.}~\bibnamefont {Andrea}}, \bibinfo {author} {\bibfnamefont
  {O.}~\bibnamefont {Marco}}, \ and\ \bibinfo {author} {\bibfnamefont
  {S.}~\bibnamefont {Alexander}},\ }\href {\doibase 10.1002/lpor.201500252}
  {\bibfield  {journal} {\bibinfo  {journal} {Laser \& Photonics Reviews}\
  }\textbf {\bibinfo {volume} {10}},\ \bibinfo {pages} {317} (\bibinfo {year}
  {2016})}\BibitemShut {NoStop}%
\bibitem [{\citenamefont {Ndagano}\ \emph {et~al.}(2017)\citenamefont
  {Ndagano}, \citenamefont {Perez-Garcia}, \citenamefont {Roux}, \citenamefont
  {McLaren}, \citenamefont {Rosales-Guzman}, \citenamefont {Zhang},
  \citenamefont {Mouane}, \citenamefont {Hernandez-Aranda}, \citenamefont
  {Konrad},\ and\ \citenamefont {Forbes}}]{Ndagano2017}%
  \BibitemOpen
  \bibfield  {author} {\bibinfo {author} {\bibfnamefont {B.}~\bibnamefont
  {Ndagano}}, \bibinfo {author} {\bibfnamefont {B.}~\bibnamefont
  {Perez-Garcia}}, \bibinfo {author} {\bibfnamefont {F.~S.}\ \bibnamefont
  {Roux}}, \bibinfo {author} {\bibfnamefont {M.}~\bibnamefont {McLaren}},
  \bibinfo {author} {\bibfnamefont {C.}~\bibnamefont {Rosales-Guzman}},
  \bibinfo {author} {\bibfnamefont {Y.}~\bibnamefont {Zhang}}, \bibinfo
  {author} {\bibfnamefont {O.}~\bibnamefont {Mouane}}, \bibinfo {author}
  {\bibfnamefont {R.~I.}\ \bibnamefont {Hernandez-Aranda}}, \bibinfo {author}
  {\bibfnamefont {T.}~\bibnamefont {Konrad}}, \ and\ \bibinfo {author}
  {\bibfnamefont {A.}~\bibnamefont {Forbes}},\ }\href
  {http://dx.doi.org/10.1038/nphys4003} {\bibfield  {journal} {\bibinfo
  {journal} {Nat Phys}\ }\textbf {\bibinfo {volume} {13}},\ \bibinfo {pages}
  {397} (\bibinfo {year} {2017})}\BibitemShut {NoStop}%
\bibitem [{\citenamefont {Amaral}\ \emph {et~al.}(2015)\citenamefont {Amaral},
  \citenamefont {Cunha},\ and\ \citenamefont {Cabello}}]{PhysRevA.92.062125}%
  \BibitemOpen
  \bibfield  {author} {\bibinfo {author} {\bibfnamefont {B.}~\bibnamefont
  {Amaral}}, \bibinfo {author} {\bibfnamefont {M.~T.}\ \bibnamefont {Cunha}}, \
  and\ \bibinfo {author} {\bibfnamefont {A.}~\bibnamefont {Cabello}},\ }\href
  {\doibase 10.1103/PhysRevA.92.062125} {\bibfield  {journal} {\bibinfo
  {journal} {Phys. Rev. A}\ }\textbf {\bibinfo {volume} {92}},\ \bibinfo
  {pages} {062125} (\bibinfo {year} {2015})}\BibitemShut {NoStop}%
\bibitem [{\citenamefont {G\"uhne}\ \emph {et~al.}(2010)\citenamefont
  {G\"uhne}, \citenamefont {Kleinmann}, \citenamefont {Cabello}, \citenamefont
  {Larsson}, \citenamefont {Kirchmair}, \citenamefont {Z\"ahringer},
  \citenamefont {Gerritsma},\ and\ \citenamefont {Roos}}]{PhysRevA.81.022121}%
  \BibitemOpen
  \bibfield  {author} {\bibinfo {author} {\bibfnamefont {O.}~\bibnamefont
  {G\"uhne}}, \bibinfo {author} {\bibfnamefont {M.}~\bibnamefont {Kleinmann}},
  \bibinfo {author} {\bibfnamefont {A.}~\bibnamefont {Cabello}}, \bibinfo
  {author} {\bibfnamefont {J.-A.}\ \bibnamefont {Larsson}}, \bibinfo {author}
  {\bibfnamefont {G.}~\bibnamefont {Kirchmair}}, \bibinfo {author}
  {\bibfnamefont {F.}~\bibnamefont {Z\"ahringer}}, \bibinfo {author}
  {\bibfnamefont {R.}~\bibnamefont {Gerritsma}}, \ and\ \bibinfo {author}
  {\bibfnamefont {C.~F.}\ \bibnamefont {Roos}},\ }\href {\doibase
  10.1103/PhysRevA.81.022121} {\bibfield  {journal} {\bibinfo  {journal} {Phys.
  Rev. A}\ }\textbf {\bibinfo {volume} {81}},\ \bibinfo {pages} {022121}
  (\bibinfo {year} {2010})}\BibitemShut {NoStop}%
\bibitem [{\citenamefont {Abramsky}\ and\ \citenamefont
  {Brandenburger}(2011)}]{1367-2630-13-11-113036}%
  \BibitemOpen
  \bibfield  {author} {\bibinfo {author} {\bibfnamefont {S.}~\bibnamefont
  {Abramsky}}\ and\ \bibinfo {author} {\bibfnamefont {A.}~\bibnamefont
  {Brandenburger}},\ }\href {http://stacks.iop.org/1367-2630/13/i=11/a=113036}
  {\bibfield  {journal} {\bibinfo  {journal} {New Journal of Physics}\ }\textbf
  {\bibinfo {volume} {13}},\ \bibinfo {pages} {113036} (\bibinfo {year}
  {2011})}\BibitemShut {NoStop}%
\bibitem [{\citenamefont {Klyachko}\ \emph {et~al.}(2008)\citenamefont
  {Klyachko}, \citenamefont {Can}, \citenamefont
  {Binicio\ifmmode~\breve{g}\else \u{g}\fi{}lu},\ and\ \citenamefont
  {Shumovsky}}]{Klyachko_2008}%
  \BibitemOpen
  \bibfield  {author} {\bibinfo {author} {\bibfnamefont {A.~A.}\ \bibnamefont
  {Klyachko}}, \bibinfo {author} {\bibfnamefont {M.~A.}\ \bibnamefont {Can}},
  \bibinfo {author} {\bibfnamefont {S.}~\bibnamefont
  {Binicio\ifmmode~\breve{g}\else \u{g}\fi{}lu}}, \ and\ \bibinfo {author}
  {\bibfnamefont {A.~S.}\ \bibnamefont {Shumovsky}},\ }\href {\doibase
  10.1103/PhysRevLett.101.020403} {\bibfield  {journal} {\bibinfo  {journal}
  {Phys. Rev. Lett.}\ }\textbf {\bibinfo {volume} {101}},\ \bibinfo {pages}
  {020403} (\bibinfo {year} {2008})}\BibitemShut {NoStop}%
\bibitem [{\citenamefont {Cabello}(2013)}]{PhysRevLett.110.060402}%
  \BibitemOpen
  \bibfield  {author} {\bibinfo {author} {\bibfnamefont {A.}~\bibnamefont
  {Cabello}},\ }\href {\doibase 10.1103/PhysRevLett.110.060402} {\bibfield
  {journal} {\bibinfo  {journal} {Phys. Rev. Lett.}\ }\textbf {\bibinfo
  {volume} {110}},\ \bibinfo {pages} {060402} (\bibinfo {year}
  {2013})}\BibitemShut {NoStop}%
\bibitem [{\citenamefont {Cabello}\ \emph {et~al.}()\citenamefont {Cabello},
  \citenamefont {Severini},\ and\ \citenamefont {Winter}}]{arXiv1010.2163}%
  \BibitemOpen
  \bibfield  {author} {\bibinfo {author} {\bibfnamefont {A.}~\bibnamefont
  {Cabello}}, \bibinfo {author} {\bibfnamefont {S.}~\bibnamefont {Severini}}, \
  and\ \bibinfo {author} {\bibfnamefont {A.}~\bibnamefont {Winter}},\
  }\href@noop {} {\ }\Eprint {http://arxiv.org/abs/arXiv:1010.2163v1}
  {arXiv:1010.2163v1} \BibitemShut {NoStop}%
\bibitem [{\citenamefont {Yan}(2013)}]{PhysRevLett.110.260406}%
  \BibitemOpen
  \bibfield  {author} {\bibinfo {author} {\bibfnamefont {B.}~\bibnamefont
  {Yan}},\ }\href {\doibase 10.1103/PhysRevLett.110.260406} {\bibfield
  {journal} {\bibinfo  {journal} {Phys. Rev. Lett.}\ }\textbf {\bibinfo
  {volume} {110}},\ \bibinfo {pages} {260406} (\bibinfo {year}
  {2013})}\BibitemShut {NoStop}%
\bibitem [{\citenamefont {Malinowski}\ \emph {et~al.}(2018)\citenamefont
  {Malinowski}, \citenamefont {Zhang}, \citenamefont {Leupold}, \citenamefont
  {Cabello}, \citenamefont {Alonso},\ and\ \citenamefont
  {Home}}]{PhysRevA.98.050102}%
  \BibitemOpen
  \bibfield  {author} {\bibinfo {author} {\bibfnamefont {M.}~\bibnamefont
  {Malinowski}}, \bibinfo {author} {\bibfnamefont {C.}~\bibnamefont {Zhang}},
  \bibinfo {author} {\bibfnamefont {F.~M.}\ \bibnamefont {Leupold}}, \bibinfo
  {author} {\bibfnamefont {A.}~\bibnamefont {Cabello}}, \bibinfo {author}
  {\bibfnamefont {J.}~\bibnamefont {Alonso}}, \ and\ \bibinfo {author}
  {\bibfnamefont {J.~P.}\ \bibnamefont {Home}},\ }\href {\doibase
  10.1103/PhysRevA.98.050102} {\bibfield  {journal} {\bibinfo  {journal} {Phys.
  Rev. A}\ }\textbf {\bibinfo {volume} {98}},\ \bibinfo {pages} {050102}
  (\bibinfo {year} {2018})}\BibitemShut {NoStop}%
\bibitem [{\citenamefont {Mosley}\ \emph {et~al.}(2008)\citenamefont {Mosley},
  \citenamefont {Lundeen}, \citenamefont {Smith}, \citenamefont {Wasylczyk},
  \citenamefont {U'Ren}, \citenamefont {Silberhorn},\ and\ \citenamefont
  {Walmsley}}]{PhysRevLett.100.133601}%
  \BibitemOpen
  \bibfield  {author} {\bibinfo {author} {\bibfnamefont {P.~J.}\ \bibnamefont
  {Mosley}}, \bibinfo {author} {\bibfnamefont {J.~S.}\ \bibnamefont {Lundeen}},
  \bibinfo {author} {\bibfnamefont {B.~J.}\ \bibnamefont {Smith}}, \bibinfo
  {author} {\bibfnamefont {P.}~\bibnamefont {Wasylczyk}}, \bibinfo {author}
  {\bibfnamefont {A.~B.}\ \bibnamefont {U'Ren}}, \bibinfo {author}
  {\bibfnamefont {C.}~\bibnamefont {Silberhorn}}, \ and\ \bibinfo {author}
  {\bibfnamefont {I.~A.}\ \bibnamefont {Walmsley}},\ }\href {\doibase
  10.1103/PhysRevLett.100.133601} {\bibfield  {journal} {\bibinfo  {journal}
  {Phys. Rev. Lett.}\ }\textbf {\bibinfo {volume} {100}},\ \bibinfo {pages}
  {133601} (\bibinfo {year} {2008})}\BibitemShut {NoStop}%
\bibitem [{Sup()}]{Supplementary}%
  \BibitemOpen
  \href@noop {} {}\bibinfo {note} {See Supplemental Material at [URL will be
  inserted by publisher] for the experimental details, the violation
  rubustness, the theoretical predictions, the measured correlation function
  $g_{i,i+1}$ and the non-classicality of the state in the Glauber--Sudarshan
  $P$ representation, which includes
  Refs.~\cite{Daryl2004,PhysRevLett.97.043602,PhysRevLett.119.050504}.}\BibitemShut
  {Stop}%
\bibitem [{\citenamefont {Vogel}\ and\ \citenamefont
  {Sperling}(2014)}]{PhysRevA.89.052302}%
  \BibitemOpen
  \bibfield  {author} {\bibinfo {author} {\bibfnamefont {W.}~\bibnamefont
  {Vogel}}\ and\ \bibinfo {author} {\bibfnamefont {J.}~\bibnamefont
  {Sperling}},\ }\href {\doibase 10.1103/PhysRevA.89.052302} {\bibfield
  {journal} {\bibinfo  {journal} {Phys. Rev. A}\ }\textbf {\bibinfo {volume}
  {89}},\ \bibinfo {pages} {052302} (\bibinfo {year} {2014})}\BibitemShut
  {NoStop}%
\bibitem [{\citenamefont {Rigovacca}\ \emph {et~al.}(2016)\citenamefont
  {Rigovacca}, \citenamefont {Di~Franco}, \citenamefont {Metcalf},
  \citenamefont {Walmsley},\ and\ \citenamefont
  {Kim}}]{PhysRevLett.117.213602}%
  \BibitemOpen
  \bibfield  {author} {\bibinfo {author} {\bibfnamefont {L.}~\bibnamefont
  {Rigovacca}}, \bibinfo {author} {\bibfnamefont {C.}~\bibnamefont
  {Di~Franco}}, \bibinfo {author} {\bibfnamefont {B.~J.}\ \bibnamefont
  {Metcalf}}, \bibinfo {author} {\bibfnamefont {I.~A.}\ \bibnamefont
  {Walmsley}}, \ and\ \bibinfo {author} {\bibfnamefont {M.~S.}\ \bibnamefont
  {Kim}},\ }\href {\doibase 10.1103/PhysRevLett.117.213602} {\bibfield
  {journal} {\bibinfo  {journal} {Phys. Rev. Lett.}\ }\textbf {\bibinfo
  {volume} {117}},\ \bibinfo {pages} {213602} (\bibinfo {year}
  {2016})}\BibitemShut {NoStop}%
\bibitem [{\citenamefont {Springer}\ \emph {et~al.}(2009)\citenamefont
  {Springer}, \citenamefont {Lee}, \citenamefont {Bellini},\ and\ \citenamefont
  {Kim}}]{PhysRevA.79.062303}%
  \BibitemOpen
  \bibfield  {author} {\bibinfo {author} {\bibfnamefont {S.~C.}\ \bibnamefont
  {Springer}}, \bibinfo {author} {\bibfnamefont {J.}~\bibnamefont {Lee}},
  \bibinfo {author} {\bibfnamefont {M.}~\bibnamefont {Bellini}}, \ and\
  \bibinfo {author} {\bibfnamefont {M.~S.}\ \bibnamefont {Kim}},\ }\href
  {\doibase 10.1103/PhysRevA.79.062303} {\bibfield  {journal} {\bibinfo
  {journal} {Phys. Rev. A}\ }\textbf {\bibinfo {volume} {79}},\ \bibinfo
  {pages} {062303} (\bibinfo {year} {2009})}\BibitemShut {NoStop}%
\bibitem [{\citenamefont {Glauber}(1963)}]{PhysRev.131.2766}%
  \BibitemOpen
  \bibfield  {author} {\bibinfo {author} {\bibfnamefont {R.~J.}\ \bibnamefont
  {Glauber}},\ }\href {\doibase 10.1103/PhysRev.131.2766} {\bibfield  {journal}
  {\bibinfo  {journal} {Phys. Rev.}\ }\textbf {\bibinfo {volume} {131}},\
  \bibinfo {pages} {2766} (\bibinfo {year} {1963})}\BibitemShut {NoStop}%
\bibitem [{\citenamefont {Sudarshan}(1963)}]{PhysRevLett.10.277}%
  \BibitemOpen
  \bibfield  {author} {\bibinfo {author} {\bibfnamefont {E.~C.~G.}\
  \bibnamefont {Sudarshan}},\ }\href {\doibase 10.1103/PhysRevLett.10.277}
  {\bibfield  {journal} {\bibinfo  {journal} {Phys. Rev. Lett.}\ }\textbf
  {\bibinfo {volume} {10}},\ \bibinfo {pages} {277} (\bibinfo {year}
  {1963})}\BibitemShut {NoStop}%
\bibitem [{\citenamefont {Leonard~Mandel}(1995)}]{book:14435}%
  \BibitemOpen
  \bibfield  {author} {\bibinfo {author} {\bibfnamefont {E.~W.}\ \bibnamefont
  {Leonard~Mandel}},\ }\href@noop {} {\emph {\bibinfo {title} {Optical
  Coherence and Quantum Optics}}},\ \bibinfo {edition} {1st}\ ed.\ (\bibinfo
  {publisher} {Cambridge University Press},\ \bibinfo {year}
  {1995})\BibitemShut {NoStop}%
\bibitem [{\citenamefont {Resch}\ \emph {et~al.}(2007)\citenamefont {Resch},
  \citenamefont {Pregnell}, \citenamefont {Prevedel}, \citenamefont
  {Gilchrist}, \citenamefont {Pryde}, \citenamefont {O'Brien},\ and\
  \citenamefont {White}}]{PhysRevLett.98.223601}%
  \BibitemOpen
  \bibfield  {author} {\bibinfo {author} {\bibfnamefont {K.~J.}\ \bibnamefont
  {Resch}}, \bibinfo {author} {\bibfnamefont {K.~L.}\ \bibnamefont {Pregnell}},
  \bibinfo {author} {\bibfnamefont {R.}~\bibnamefont {Prevedel}}, \bibinfo
  {author} {\bibfnamefont {A.}~\bibnamefont {Gilchrist}}, \bibinfo {author}
  {\bibfnamefont {G.~J.}\ \bibnamefont {Pryde}}, \bibinfo {author}
  {\bibfnamefont {J.~L.}\ \bibnamefont {O'Brien}}, \ and\ \bibinfo {author}
  {\bibfnamefont {A.~G.}\ \bibnamefont {White}},\ }\href {\doibase
  10.1103/PhysRevLett.98.223601} {\bibfield  {journal} {\bibinfo  {journal}
  {Phys. Rev. Lett.}\ }\textbf {\bibinfo {volume} {98}},\ \bibinfo {pages}
  {223601} (\bibinfo {year} {2007})}\BibitemShut {NoStop}%
\bibitem [{\citenamefont {Thomas-Peter}\ \emph {et~al.}(2011)\citenamefont
  {Thomas-Peter}, \citenamefont {Smith}, \citenamefont {Datta}, \citenamefont
  {Zhang}, \citenamefont {Dorner},\ and\ \citenamefont
  {Walmsley}}]{PhysRevLett.107.113603}%
  \BibitemOpen
  \bibfield  {author} {\bibinfo {author} {\bibfnamefont {N.}~\bibnamefont
  {Thomas-Peter}}, \bibinfo {author} {\bibfnamefont {B.~J.}\ \bibnamefont
  {Smith}}, \bibinfo {author} {\bibfnamefont {A.}~\bibnamefont {Datta}},
  \bibinfo {author} {\bibfnamefont {L.}~\bibnamefont {Zhang}}, \bibinfo
  {author} {\bibfnamefont {U.}~\bibnamefont {Dorner}}, \ and\ \bibinfo {author}
  {\bibfnamefont {I.~A.}\ \bibnamefont {Walmsley}},\ }\href {\doibase
  10.1103/PhysRevLett.107.113603} {\bibfield  {journal} {\bibinfo  {journal}
  {Phys. Rev. Lett.}\ }\textbf {\bibinfo {volume} {107}},\ \bibinfo {pages}
  {113603} (\bibinfo {year} {2011})}\BibitemShut {NoStop}%
\bibitem [{\citenamefont {Kleinmann}\ \emph {et~al.}(2011)\citenamefont
  {Kleinmann}, \citenamefont {Gühne}, \citenamefont {Portillo}, \citenamefont
  {Åke Larsson},\ and\ \citenamefont {Cabello}}]{1367-2630-13-11-113011}%
  \BibitemOpen
  \bibfield  {author} {\bibinfo {author} {\bibfnamefont {M.}~\bibnamefont
  {Kleinmann}}, \bibinfo {author} {\bibfnamefont {O.}~\bibnamefont {Gühne}},
  \bibinfo {author} {\bibfnamefont {J.~R.}\ \bibnamefont {Portillo}}, \bibinfo
  {author} {\bibfnamefont {J.}~\bibnamefont {Åke Larsson}}, \ and\ \bibinfo
  {author} {\bibfnamefont {A.}~\bibnamefont {Cabello}},\ }\href
  {http://stacks.iop.org/1367-2630/13/i=11/a=113011} {\bibfield  {journal}
  {\bibinfo  {journal} {New Journal of Physics}\ }\textbf {\bibinfo {volume}
  {13}},\ \bibinfo {pages} {113011} (\bibinfo {year} {2011})}\BibitemShut
  {NoStop}%
\bibitem [{\citenamefont {Cabello}\ \emph {et~al.}(2018)\citenamefont
  {Cabello}, \citenamefont {Gu}, \citenamefont {G\"uhne},\ and\ \citenamefont
  {Xu}}]{PhysRevLett.120.130401}%
  \BibitemOpen
  \bibfield  {author} {\bibinfo {author} {\bibfnamefont {A.}~\bibnamefont
  {Cabello}}, \bibinfo {author} {\bibfnamefont {M.}~\bibnamefont {Gu}},
  \bibinfo {author} {\bibfnamefont {O.}~\bibnamefont {G\"uhne}}, \ and\
  \bibinfo {author} {\bibfnamefont {Z.-P.}\ \bibnamefont {Xu}},\ }\href
  {\doibase 10.1103/PhysRevLett.120.130401} {\bibfield  {journal} {\bibinfo
  {journal} {Phys. Rev. Lett.}\ }\textbf {\bibinfo {volume} {120}},\ \bibinfo
  {pages} {130401} (\bibinfo {year} {2018})}\BibitemShut {NoStop}%
\bibitem [{\citenamefont {Achilles}\ \emph {et~al.}(2004)\citenamefont
  {Achilles}, \citenamefont {Silberhorn}, \citenamefont {Sliwa}, \citenamefont
  {Banaszek}, \citenamefont {Walmsley}, \citenamefont {Fitch}, \citenamefont
  {Jacobs}, \citenamefont {Pittman},\ and\ \citenamefont
  {Franson}}]{Daryl2004}%
  \BibitemOpen
  \bibfield  {author} {\bibinfo {author} {\bibfnamefont {D.}~\bibnamefont
  {Achilles}}, \bibinfo {author} {\bibfnamefont {C.}~\bibnamefont
  {Silberhorn}}, \bibinfo {author} {\bibfnamefont {C.}~\bibnamefont {Sliwa}},
  \bibinfo {author} {\bibfnamefont {K.}~\bibnamefont {Banaszek}}, \bibinfo
  {author} {\bibfnamefont {I.~A.}\ \bibnamefont {Walmsley}}, \bibinfo {author}
  {\bibfnamefont {M.~J.}\ \bibnamefont {Fitch}}, \bibinfo {author}
  {\bibfnamefont {B.~C.}\ \bibnamefont {Jacobs}}, \bibinfo {author}
  {\bibfnamefont {T.~B.}\ \bibnamefont {Pittman}}, \ and\ \bibinfo {author}
  {\bibfnamefont {J.~D.}\ \bibnamefont {Franson}},\ }\href {\doibase
  10.1080/09500340408235288} {\bibfield  {journal} {\bibinfo  {journal}
  {Journal of Modern Optics}\ }\textbf {\bibinfo {volume} {51}},\ \bibinfo
  {pages} {1499} (\bibinfo {year} {2004})}\BibitemShut {NoStop}%
\bibitem [{\citenamefont {Achilles}\ \emph {et~al.}(2006)\citenamefont
  {Achilles}, \citenamefont {Silberhorn},\ and\ \citenamefont
  {Walmsley}}]{PhysRevLett.97.043602}%
  \BibitemOpen
  \bibfield  {author} {\bibinfo {author} {\bibfnamefont {D.}~\bibnamefont
  {Achilles}}, \bibinfo {author} {\bibfnamefont {C.}~\bibnamefont
  {Silberhorn}}, \ and\ \bibinfo {author} {\bibfnamefont {I.~A.}\ \bibnamefont
  {Walmsley}},\ }\href {\doibase 10.1103/PhysRevLett.97.043602} {\bibfield
  {journal} {\bibinfo  {journal} {Phys. Rev. Lett.}\ }\textbf {\bibinfo
  {volume} {97}},\ \bibinfo {pages} {043602} (\bibinfo {year}
  {2006})}\BibitemShut {NoStop}%
\bibitem [{\citenamefont {Abramsky}\ \emph {et~al.}(2017)\citenamefont
  {Abramsky}, \citenamefont {Barbosa},\ and\ \citenamefont
  {Mansfield}}]{PhysRevLett.119.050504}%
  \BibitemOpen
  \bibfield  {author} {\bibinfo {author} {\bibfnamefont {S.}~\bibnamefont
  {Abramsky}}, \bibinfo {author} {\bibfnamefont {R.~S.}\ \bibnamefont
  {Barbosa}}, \ and\ \bibinfo {author} {\bibfnamefont {S.}~\bibnamefont
  {Mansfield}},\ }\href {\doibase 10.1103/PhysRevLett.119.050504} {\bibfield
  {journal} {\bibinfo  {journal} {Phys. Rev. Lett.}\ }\textbf {\bibinfo
  {volume} {119}},\ \bibinfo {pages} {050504} (\bibinfo {year}
  {2017})}\BibitemShut {NoStop}%
\end{thebibliography}
\end{document}